\pdfoutput=1
\documentclass{article}

\usepackage[preprint]{neurips_2024}

\usepackage[utf8]{inputenc} % allow utf-8 input
\usepackage[T1]{fontenc}    % use 8-bit T1 fonts
\usepackage{hyperref}       % hyperlinks
\usepackage{url}            % simple URL typesetting
\usepackage{booktabs}       % professional-quality tables
\usepackage{amsfonts}       % blackboard math symbols
\usepackage{nicefrac}       % compact symbols for 1/2, etc.
\usepackage{microtype}      % microtypography
\usepackage{xcolor}         % colors

\newcommand{\etal}{\textit{et al.}}
\usepackage{comment}
\usepackage{amsthm}
\usepackage{float}
\usepackage{graphicx}
\usepackage{subfigure}
\usepackage{caption}
\usepackage{amsmath,amssymb}
\usepackage{mathtools}
\usepackage{booktabs}
\usepackage{multirow}
\usepackage{siunitx}
\usepackage{adjustbox}
\usepackage{pifont}
\usepackage{scalerel}
\usepackage{tabularray}
\usepackage{makecell}
\usepackage{wrapfig}

\title{Federated Domain-Specific Knowledge Transfer on Large Language Models Using Synthetic Data }

\author{{\bf Haoran Li$^*$}$^1$, {\bf Xinyuan Zhao$^*$}$^{2}$, {\bf Dadi Guo$^*$}$^{3}$, {\bf Hanlin Gu$\footnotemark[4]$} \ $^{4}$\\
 {\bf Ziqian Zeng}$^{5}$,  {\bf Yuxing Han}$^{2}$, {\bf Yangqiu Song}$^{1}$, {\bf Lixin Fan}$^{4}$, {\bf Qiang Yang}$^{1, 4}$\\
$^{1}$The Hong Kong University of Science and Technology\\
$^{2}$Shenzhen International Graduate School, Tsinghua University, Shenzhen 518055, China \\
$^{3}$Center for Data Science, AAIS, Peking University
$^{4}$WeBank, China  \\
$^{5}$South China University of Technology\\
\texttt{hlibt@connect.ust.hk}
}

\begin{document}
\theoremstyle{definition}
\newtheorem{definition}{Definition}[section]

\maketitle
\renewcommand{\thefootnote}{\fnsymbol{footnote}}
\footnotetext[4]{Corresponding author.}

\begin{abstract}
% As large language models (LLMs) demonstrate unparalleled performance and generalization ability, LLMs are widely used and integrated into various applications.
% Yet, when it comes to federated learning scenarios on sensitive domains, most clients refuse to use external LLMs even though they do have incentives to obtain relevant knowledge from LLMs to improve their own task performance.
% On one hand, strict regulations on data security and privacy forbid these clients from disclosing sensitive or private data to external LLMs without formal authorization.
% On the other hand, with limited computational resources and task-specific data, clients can only afford to train small language models (LMs). 
% Based on the common observation that LLMs can empower small LMs, in this paper, we present \lhr{\underline{G}ener\underline{A}tive Fede\underline{RA}ted Knowled\underline{G}e \underline{E}xtraction (GARAGE)} framework to selectively extract knowledge from LLMs via data augmentation and knowledge distillation in a generative and privacy-preserving manner.
% Compared with the previous approaches on privacy-preserving knowledge distillation, our proposed \lhr{GARAGE} can selectively distill the knowledge desired by the clients while preserving clients' data privacy.
% Moreover, \lhr{GARAGE} can be flexibly applied across various LLMs' architectures with no limitations.
% We conduct extensive experiments to show that \lhr{GARAGE} is able to improve \lhr{XXXX} on locally trained LMs.

As large language models (LLMs) demonstrate unparalleled performance and generalization ability, LLMs are widely used and integrated into various applications. 
When it comes to sensitive domains, as commonly described in federated learning scenarios, directly using external LLMs on private data is strictly prohibited by stringent data security and privacy regulations.
For local clients, the utilization of LLMs to improve the domain-specific small language models (SLMs), characterized by limited computational resources and domain-specific data, has attracted considerable research attention. 
By observing that LLMs can empower domain-specific SLMs, existing methods predominantly concentrate on leveraging the public data or LLMs to generate more data to transfer knowledge from LLMs to SLMs. 
However, due to the discrepancies between LLMs' generated data and clients' domain-specific data, these methods cannot yield substantial improvements in the domain-specific tasks.
In this paper, we introduce a Federated Domain-specific Knowledge Transfer (FDKT) framework, which enables domain-specific knowledge transfer from LLMs to SLMs while preserving clients’ data privacy. 
The core insight is to leverage LLMs to augment data based on domain-specific few-shot demonstrations, which are synthesized from private domain data using differential privacy.
Such synthetic samples share similar data distribution with clients' private data and allow the server LLM to generate particular knowledge to improve clients' SLMs.
% Specifically, clients firstly generate the domain-specific synthetic data with differential privacy and upload it to the server. 
% Then, LLMs augment the synthetic data to transfer their relevant knowledge to help improve the domain-specific SLMs. 
% \lhr{Compared with the previous approaches on privacy-preserving knowledge transfer, our proposed OFKT can selectively distill the knowledge desired by the clients while preserving clients' data privacy.
% }
The extensive experimental results demonstrate that the proposed FDKT framework consistently and greatly improves SLMs' task performance by around 5\% with a privacy budget of less than 10, compared to local training on private data.
%We will release the reproducible code.%~\footnote{Our code is available at \url{1}}.

% Moreover, OFKT can be flexibly applied across various LLMs' architectures without limitation. We conduct extensive experiments to show that OFKT is able to improve \lhr{XXXX} on locally trained LMs.

\end{abstract}

\section{Introduction}
\label{1-intro}
%%% selective distill
%%% 
Presently, the generative large language models (LLMs) are revolutionizing the existing paradigms in Natural Language Processing (NLP) tasks into a generation pipeline~\cite{2020t5, Brown2020LanguageMA, OpenAI2023GPT4TR, ouyang2022training}. 
With the support of extensive training data and careful fine-tuning, LLMs exhibit unparalleled capabilities in comprehension and adherence to instructions, reasoning~\cite{Kojima2022LargeLM, Wei2022ChainOT}, planning~\cite{yuan-etal-2023-distilling, pmlr-v162-huang22a} and generalization to unseen tasks~\cite{Chen2021EvaluatingLL, zhou2023leasttomost,sanh2022multitask, 2022flant5}.
Hence, both research and engineering efforts are made to build LLM-empowered autonomous systems~\cite{Significant_Gravitas_AutoGPT, Park-2023-Generative, liang2023taskmatrix, Shen2023HuggingGPTSA} to exploit LLMs as agents for complex tasks.
However, for sensitive applications that emphasize the protection of data security and privacy, external LLMs yet cannot be directly utilized due to their inherent privacy vulnerabilities~\cite{li-etal-2023-multi-step, Greshake-2023-not}.

To protect data privacy on sensitive domains, federated learning (FL)~\cite{YangLCT19, mcmahan2017communication, konevcny2016federated} has been proposed to collaboratively build machine learning models without compromising on clients' data privacy.
Conventionally, for FL, clients (the data holders) train their models locally and optimize their model weights according to all clients' aggregated model weights (FedAvg)~\cite{konevcny2016federated,konevcny2016federated2} coordinated by the server.
When it comes to federated LLMs, a few recent works~\cite{wang2023can, deng2023mutual, ye2024openfedllm,kang2023grounding} considered data holders with domain-specific small LMs (SLMs), i.e., as the clients and LLMs' service providers as the servers. 
In addition to aggregating knowledge from other clients, each client can directly learn from the server LLM via knowledge distillation to improve its local SLMs' performance.

This formulation presents an unresolved challenge for clients: their personalized private data, denoted as $D \sim \mathcal{D}$, which follows an unknown distribution $\mathcal{D}$, is often limited in quantity. 
A naive solution is to upload the private domain-specific data to the server and allow the LLMs to augment more data which follows $\mathcal{D}$. 
However, this approach is not viable due to privacy constraints.
Therefore, a series of works aims to improve SLMs with the aid of LLMs without disclosing $D$. 
The knowledge distillation method~\cite{wang2023can} transfers the knowledge from LLMs to SLMs based on the public data $D_p$. 
Nevertheless, there is a discrepancy between $D_p$ and $D$ because $D_p$ may not necessarily follow the distribution $\mathcal{D}$. 
This difference prevents these methods from effectively improving the SLMs.
Other augmentation methods \cite{deng2023mutual}  mitigate this discrepancy by utilizing LLMs to generate data according to private labels. 
Still, the augmented data causes a misalignment with the actual distribution $\mathcal{D}$.

To address the aforementioned limitations, in this work, we propose a Federated Domain-specific Knowledge Transfer (FDKT) framework.
%\lhr{FDKT implements a pure generative pipeline for the client SLM to selectively learn knowledge from the server LLM with differential privacy (DP)~\cite{Dwork-08-DP} guarantee. }
FDKT implements a generative pipeline on private data $D$ by leveraging LLMs to augment the data according to domain-specific examples. 
These domain-specific examples are generated from the private data distribution $\mathcal{D}$ with differential privacy (DP)~\cite{Dwork-08-DP} guarantee, resulting in synthetic data. 
Due to the introduction of DP's noise, the synthetic data may contain artifacts. 
To address this discrepancy, we further exploit the
sever LLM for clustering-based filtering and augmentation to correct the artifacts.
The contributions of our proposed FDKT are summarized below:
%Specifically, to transfer knowledge to domain-specific SLMs while protecting private local data, FDKT first generates differentially private synthetic data $\mathcal D'$ with respect to the client's private data.
%Then, the synthetic data is sent to the server LLM for both in-context data augmentation and careful data filtering procedures. Finally, the server only needs to return the augmented data $\mathcal D^a$ to train clients' SLMs to solve clients' specific tasks.

\begin{itemize}

\item 
FDKT enables domain-specific knowledge transfer from LLMs to SLMs. The client transmits synthetic data conditionally generated on its private data to glean required knowledge from the server-side LLM. The server can then impart the client's domain-oriented knowledge to improve each client's customized task performance.

\item FDKT prioritizes privacy. 
To protect the privacy of clients' sensitive data, FDKT minimizes potential threats by sharing synthetic and differentially private data to the server.
Simultaneously, to protect the server's intellectual property, FDKT only requires API-level access to the server LLM without exposing any unnecessary hidden information.

\item FDKT is versatile across various model architectures for both the server-side LLM and client-side SLM, hence ensuring comprehensive applicability.

\item Experimental results demonstrate that our proposed FDKT can consistently improve individual client SLM's accuracies significantly.
Moreover, FDKT effectively facilitates multi-task learning across multiple clients for the one-to-many scenario.
%Without training on the private local data, LMs fine-tuned with knowledge extracted by FDKT already surpass baseline LMs trained on private local data.

%%% leave content for in domain & out domain 
\end{itemize}

%Wang \etal~\cite{wang2023can}
%\section{Related Works}
%\label{relate}

%\textbf{Federated Learning on LLMs}.

%\textbf{Differential}

%\textbf{Synthetic Data Generation with DP}. 
\section{Preliminaries}
\label{sec: pre}

\subsection{Federated Learning on LLMs}

Adapting general-purpose LLMs~\cite{yang2023harnessing,zhou2023comprehensive, OpenAI2023GPT4TR,touvron2023llama} to downstream tasks typically involves the full fine-tuning of all model parameters. 
However, this approach can be prohibitively expensive, especially for domain-specific tasks. 
To mitigate this challenge, Parameter-Efficient Fine-Tuning (PEFT) methods \cite{houlsby2019parameter,he2021towards,lester2021power,li2021prefix,hu2021lora} have been proposed. 
%These methods efficiently adapt LLMs to specific domains or tasks by fine-tuning only a subset of model parameters while keeping the rest frozen. 
PEFT methods provide a direct solution to the challenges of communication overhead and fine-tuning costs in federated learning for large language models~\cite{collins2023profit,babakniya2023slora}. Several studies have extended PEFT methods in the context of FL for LLMs, including FedPETuning~\cite{zhang2022federated}, Federated Adapter Tuning~\cite{cai2022autofednlp}, Federated Prompt Tuning \cite{zhao2022reduce} and FATE-LLM \cite{fan2023fate}. Specifically, the FedPETuning methods proposed by \cite{zhang2022federated} have demonstrated a significant reduction in communication overhead in the FL setting. 
Additionally, they found that PEFT methods can effectively reduce local model adaptation costs for clients in FL systems. 
%These findings suggest that FL clients, such as devices with limited storage capacity, can greatly benefit from PEFT methods. 
These methods enable the sharing of LLMs across different tasks while maintaining only a few parameters for each task, thereby reducing the storage requirement. 
%By leveraging PEFT methods, FL clients can efficiently adapt LLMs to their specific needs while minimizing communication overhead and fine-tuning costs.

%%% hr add bo li and yang liu's papers for related works
%wang2023can
%deng2023mutual
In addition to PEFT methods, a few recent works~\cite{wang2023can, deng2023mutual} explore the transfer learning approach to transfer server LLMs' knowledge into client SLMs.
Wang \etal \cite{wang2023can} propose knowledge distillation based on publicly available data while Deng \etal \cite{deng2023mutual} expliot the server LLM to augment data via prompting with general descriptions about domain and label information.
However, these works focus on the general knowledge transfer pipeline and fail to exploit rich domain characteristics inside clients' private data due to privacy considerations.
In contrast, we propose to transfer the server LLM's knowledge using domain-oriented and privacy-preserving synthetic data that share a similar distribution with clients' private data.

\subsection{Differential Privacy}
In this section, we introduce the formal definition of Differential Privacy (DP)~\cite{Dwork-08-DP}:
\begin{definition}[Differential Privacy]
\label{def:dp}
A randomized \textit{mechanism} $M$ with domain $\mathcal{X}$ and range $\mathcal{R}$ satisfies $(\epsilon,\delta)$-\textit{differential privacy} if for any two neighboring datasets $D_1, D_2$ that only differ in one element and for any subsets of output $O \subseteq \mathcal{R}$:
\begin{equation}\label{eq:dp-bound}
%\small
Pr[M(D_1) \in O] \leq e^{\epsilon}Pr[M(D_2)\in O]+\delta.
\end{equation}
\end{definition}
The definition provided by DP introduces the concept of  \textit{plausible deniability}~\cite{Bindschaedler-2017-Plausible} and establishes bounded privacy parameters $(\epsilon, \delta)$ that serve to quantify the effectiveness of the mechanisms under scrutiny.
Regarding deep learning models, DPSGD~\cite{Abadi2016DeepLW} injects Gaussian noise into the models' gradients so that the trained models are differentially private with respect to their training data.
In addition, according to the Post-Processing Theorem~\cite{Dwork-08-DP}, for any mapping $g$, the post-processing $g \circ M$ is also $(\epsilon, \delta)$-DP.
Thus, the trained models can be safely released for public usage.

\subsection{DP-tuned LMs}
To enhance data privacy within LMs, a majority of research focusing on privacy-preserving LMs primarily incorporates DPSGD as the foundational component.
DPSGD's usage can be summarized into four parts.
The first part is DP fine-tuning that fine-tunes LMs on sensitive downstream datasets~\cite{Feyisetan-20-Privacy, qu2021natural, shi-etal-2022-just, yu2022differentially,li2023p, li2022large, huang2023privacy, yu2023selective}.
Though DP fine-tuning can achieve comparable performance as normal fine-tuning on several NLP tasks, it is time-consuming to train on the downstream datasets.
Hence, the second part proposes DP pre-training to pre-train privacy-preserving LMs so that no more fine-tuning is needed for downstream tasks.
DP-BART~\cite{Igamberdiev-2023-DP-BART} considered text rewriting under LDP~\cite{kasiviswanathan-2011-localdp} to rewrite the input with DP guarantee.
The third part focuses on generative LLMs and proposes various DP-based prompt-tuning methods~\cite{Ozdayi2023, li2023privacy, duan2023flocks, hong2024dpopt} which leverage prompt tricks to protect privacy during LLM interactions.
Lastly, the fourth part proposes DP-based synthetic text generation to conditionally sample text from DP-tuned generative LMs~\cite{yue2022synthetic, Kurakin2023HarnessingLM, mattern-2022-differentially, flemings2024differentially}.
Their experimental findings indicate that language models fine-tuned with synthetic texts can outperform LMs that have been tuned directly with DPSGD in terms of testing performance. 
In our approach, we suggest enabling clients to share DP-sanitized synthetic texts with the server, thereby facilitating the transfer of client-specific knowledge without compromising data privacy.

\section{Federated Domain-Specific Knowledge Transfer}
\label{sec: method}

\begin{figure*}[t]
\centering
\includegraphics[width=0.998\textwidth]{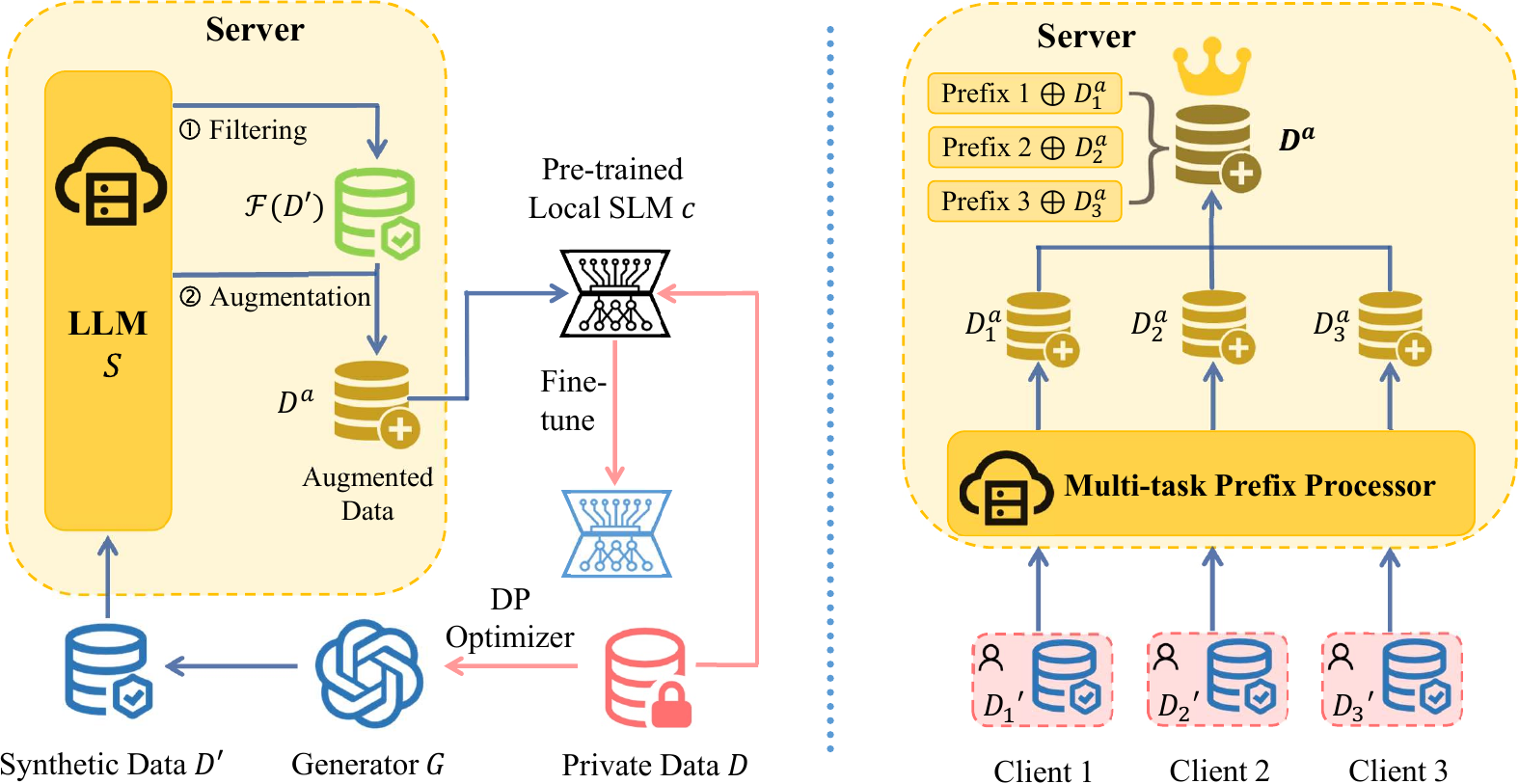}
%\vspace{-0.1in}
\caption{
Overview of FDKT's selective knowledge transfer pipeline.
The left subfigure illustrates the workflow of FDKT for enhancing individual client performance, while the right subpart depicts how FDKT facilitates federated training across multiple clients for multi-task learning.
The yellow region is under the control of the server and the rest part belongs to the client.
%The rose lines indicate the normal local fine-tuning pipeline that directly fine-tunes $c$ on private data $D$.
The rose lines involve interactions with private data $D$.
In contrast, blue lines represent interactions that do not disclose $D$.
In all interactions between FDKT's client and server, only synthetic data $D'$ is exchanged to facilitate knowledge transfer.
In the right subpart, the multi-task prefix processor adds task-dependent prefixes to each client's augmented data to train multi-task SLMs.
%Client LM $c$ fine-tuned on $D^r$ and $D^a$ outperforms local fine-tuning for the client's own downstream tasks.
}
\label{fig:garage}
\vspace{-0.19in}
\end{figure*}

In this section, we present the detailed workflow of our proposed Federated Domain-specific Knowledge Transfer (FDKT).
First, we formulate the problem for the 1-to-1 server-client setting based on their capabilities and incentives.
Then, from the client's perspective, we show how synthetic and privacy-preserving data are generated.
Next, for the server side, 
we introduce in-context data augmentation with careful selection mechanisms of the augmented data to generate the client's required knowledge.
Finally, we extend FDKT to handle multiple clients and train multi-task SLMs across multiple sensitive domains collaboratively.
Figure~\ref{fig:garage} depicts FDKT's whole workflow.

%\newpage
% \subsection{Threat Model}
% \HL{Consider two parties involving the server who owns a LLM $f_\theta$ parameterized by $\theta$ and one client that has a SLM $g_{\omega}$ parameterized by ${\omega}$ and private data $D_p$. In this work, we consider a semi-honest server, who may faithfully executes the training protocol
% but may launch privacy attacks to infer the private data of the clients.}

\subsection{Problem Formulation}
% \HL{This paper consider the client aims to improve his model performance by leveraging the LLMs. Specifically, the server use the data $D_g$ generated by LLM helps the SLM of the client to train $g_{\omega}$, i.e.,  
% \begin{equation}
%     \ell(\omega; D_g),
% \end{equation} 
% where $\ell$ is the...Some existing methods \cite{} used the public dataset as $D_g$, however, they ignored the difference between the public and private dataset. Therefore, a direct idea is let the data $D_g$ generated based on private data $D_p$ such that the $D_g$ has some similar domain information as the client as the following:
% \begin{equation}
%    \ell(\omega;  D_g|D_p).  
% \end{equation}
% However, there exists one challenge that the private data $D_p$ of the client cannot be transferred to the server. Therefore, how to look for a privacy-preserving mechanishm $P()$ to help the LLM generated the useful for distilling the SLM. In summarize, the problem formulation is defined as:
% \begin{equation}
%      \ell(\omega;  D_g|P(D_p)).  
% \end{equation}
% }

This paper takes both the client's and server's incentives into consideration. 
The client's goal is to improve its SLM's performance by leveraging the server LLM.
From the server's perspective, the server is also reluctant to transfer excessive knowledge or reveal its LLM's hidden aspects to safeguard intellectual property.
Without loss of generality, we start from the one-to-one configuration where there is a server with only one client and we incorporate FDKT to improve the client SLM's performance individually.
%In other words, FDKT offers SLMs a better initialization strategy to selectively acquire knowledge from the server to improve the client LM's domain-specific performance.
We assume the client possesses a private local dataset $D = {(x_i, y_i)}_{i=1}^{N}$ where $N$ is relatively small and has limited computational resources that can only operate a small-scale LM $c$ in-house.
Furthermore, the server owns a powerful LLM $S$.
Since $N$ is small, directly fine-tuning $c$ on $D$ cannot yield satisfactory results. The \textbf{threat model} we consider is semi-honest server $S$ aims to infer the private data of client $c$.

% The client's goal is to leverage the server LLM $S$'s knowledge to improve $c$'s SLM on its testing data $T$ that shares a similar distribution with $D$.
%From the server's perspective, the server has the incentive to generate the client's required knowledge for revenue.
%Concurrently, the server is reluctant to expose unnecessary or excessive information for $S$'s model security.
%Therefore, the server should balance between knowledge transfer and the protection of its proprietary model.

%\subsection{Client-side Synthetic Data Generation}

\subsection{Client-side Synthetic Data Generation}
\label{sec: syntheic}
To acquire domain-specific knowledge for its own task, the client needs to share its task-dependent data with $S$ at first for further knowledge transfer. %while avoiding disclosing any data samples in its private local data $D$. 
However, directly transmitting private local data $D$ to the server violates the client's privacy requirement and sharing existing public data cannot acquire $c$'s desired knowledge.
Inspired by recent progress on differentially private synthetic text generation~\cite{yue2022synthetic, Kurakin2023HarnessingLM}, we propose to share such synthetic data that are distributed similarly with $D$ to the server.
Specifically, we use private data $D$ to fine-tune a pre-trained generative LM with DPSGD. After fine-tuning, we obtain a differentially private generator $G$.
Finally, we can conditionally sample from $G$ to acquire the domain-specific synthetic data.

During fine-tuning, for any $(x, y) \in D$, we concatenate the data pair with task-specific prompts into a coherent string $s = $``\{$p_1$\} + \{$y$\} + \{$p_2$\} + \{$x$\}'' where $p_1$, $p_2$ are the prompts and `+' denotes the textual concatenation.
For review sentiment classification, given its review $x$ and corresponding rating $y$, we may construct the string as $s = $``Rating: \{$y$\}, Review:\texttt{\textbackslash n} \{$x$\}'' for fine-tuning.
Then, the language modeling objective is applied with teacher forcing \cite{Williams-1989-teacher} to fine-tune $G$ based on $s$:
\begin{equation}
\label{eqn:LM}
%\small
L_{G} (s;\theta_{G}) =
-\sum\nolimits_{i=1}^{u-1} \log(\text{Pr}(w_i|w_0,w_1,...,w_{i-1})),
\end{equation}
where $s=$``$w_0 w_1 ... w_{u-1}$'' is the reformatted string from $(x, y) \in D$ and $\theta_{G}$ is optimized via noise-injected DP optimizers.
In addition, the client has the discretion to determine its privacy budget parameters $(\epsilon,\delta)$ according to its specific privacy requirements.
After fine-tuning $G$, $G$'s outputs are guaranteed to be $(\epsilon,\delta)$-DP with respect to the client's private data $D$.%\textit{differential privacy}

To generate synthetic data, we adopt sampling-based decoding algorithms to conditionally generate synthetic sample $x'$ given the label $y$.
We repeatedly prompt the generator $G$ with the concatenated string consisting of ``\{$p_1$\} + \{$y$\} + \{$p_2$\}'' to generate $x'$.
By sampling, we may obtain as many synthetic $x'$ as we want via decoding multiple times.

After generating sufficient synthetic data, we obtain the synthetic dataset $D' = \{(x_i', y_i)\}_{i=1}^{|D'|}$.
Then, $D'$ can be safely shared to the server side with the DP guarantee.
Such DP-sampled synthetic data have two advantages.
First, these conditionally generated synthetic data encompass data distribution similar to that of private data.
Hence, knowledge extraction on conditionally generated $D'$ can selectively transfer the client's task-specific knowledge.
Second, based on the aforementioned Post-Processing Theorem, synthetic data are sampled from DP-tuned LMs so that the same DP guarantee used for fine-tuning is satisfied.
%On the one hand, synthetic data encompass data distribution similar to that of private data.
%On the other hand, synthetic data are sampled from DP-tuned LMs so that a strict DP guarantee is satisfied.

\subsection{Sever-side Knowledge Transfer}
To transfer the client's required knowledge, FDKT implements a generative pipeline for data augmentation with careful data selection procedures.

\subsubsection{High-quality Data Filtering Mechanism}
\label{sec: selection}

Even though conditionally generated synthetic data have already been adopted, there are still two weaknesses.
First, the quality of synthetic data often deteriorates due to the incorporation of random noise throughout the optimization process of generator $G$.
Second, synthetic data generated from the same prompt tends to exhibit similar semantics, resulting in a significant lack of diversity.
Consequently, under a similar distribution, poor-quality samples can be detrimental to data augmentation and model training. 
In light of the above observations, we propose a simple yet effective data filtering mechanism $\mathcal{F}(\cdot)$ to discard low-quality samples within the same distribution.
Our filtering mechanism includes a clustering stage and a selection stage.

In the initial stage, we compute sentence embeddings for all sentences using pre-trained sentence transformers~\cite{reimers-2019-sentence-bert}.
Subsequently, we apply K-means clustering~\cite{Hartigan-clustering} to these embeddings to group similar sentences.
Based on the number of sentences, we manually select the appropriate cluster number so that all samples within the same cluster fit within the server LLM's context length.

For the selection stage, we exploit the LLM $S$ as an evaluator~\cite{zheng2023judging,fu2023gptscore,li2024leveraging} to select high-quality samples within each cluster.
Specifically, among each cluster, we design a multiple-choice prompt template that presents the samples as options. 
We then instruct the server LLM $S$ to select half of the samples with higher quality and filter out the remaining half.
We use $\mathcal{F}(D')$ to denote the selected synthetic data after the filtering mechanism.

\subsubsection{In-context Data Augmentation}
%Even though we may sample sufficient synthetic data from $G$ as illustrated in Section~\ref{sec: syntheic}, the quality of the synthetic data may be compromised by the introduction of noise during $G$'s optimization process.
Currently, LLMs are extensively employed for in-context data augmentation, and the superior quality of the augmented data has been thoroughly investigated~\cite{wang-etal-2023-self-instruct, alpaca, sun2023principledriven, west-etal-2022-symbolic}.
Therefore, we propose to randomly sample a few selected synthetic data points from $\mathcal{F}(D')$ as demonstrations and exploit the server-side LLM $S$ to generate similar samples of better quality to further rectify errors introduced by random noise from $G$.
To conduct in-context data augmentation, we first need to prepare the augmentation prompt $I$ consisting of the task instruction and few-shot demonstrations sampled from the filtered synthetic data $\mathcal{F}(D')$ as described in Sec~\ref{sec: selection}.
Then, the augmented data can be represented as: 
\begin{equation}
\label{eqn:aug}
%\small
D^a = \{(x, y) | x  \sim S (x | I), y \sim S (y | x, I)\},
\end{equation}
where $S (x | I)$ denotes that $x$ is generated from $S$ conditioned on the augmentation prompt $I$.
By utilizing few-shot demonstrations, the server LLM $S$ can perform in-context learning to augment new data points based on its knowledge.
Such newly augmented data, $D^a$, 
not only enhance the diversity of the private dataset $D$ while preserving a similar distribution but also maintain better data quality than synthetic data since LLM $S$ is more capable than the generator $G$. 
Consequently, $D^a$ can effectively improve the generalization abilities of client SLMs.
%Additionally, since augmented data $D^a$ has similar formats as the synthetic data generated by $G$, rationale distillation in Section~\ref{sec: distill} can also be applied to $D^a$ to generate extra knowledge to $c$.
%We use $D^r_a$ to denote step-by-step distilled data from the augmented data $D^a$:
% \begin{equation}
% \label{eqn:aug+rat}
% %\small
% D^r_a = \{\big((p_r + x, r), (p_l + x, y)\big) | (x, y) \in D^a  \}.
% \end{equation}

\subsection{Local SLM Fine-tuning}
After the server sends back its augmented dataset $D^a$, we can integrate augmented pairs $(x, y)$ with private data.
During training, we can directly apply language modeling objective to fine-tune the client SLM $c$ to maximize $c$'s conditional generation probability $\text{Pr}(y | x)$ similar as Equation~\ref{eqn:LM}.
During inference, for any given query $x$, we use greedy decoding to decode the SLM $c$'s response.
% In the augmented dataset $D^a$ and the synthetic dataset $D'$, every sample is already of the $(x, y)$ format and can be readily suitable for use in language modeling.
% For the distilled datasets, every sample of $D^r$ and $D^r_a$ samples are structured as nested pairs $( (p_r + x, r), (p_l + x, y) )$.
% These can be separated into two distinct pairs $(p_r + x, r)$ and $(p_l + x, y)$ for conditional generation.
% When $p_r + x$ is prompted, the client LM $c$ aims to generate the rationale or explanation for inferring $y$.
% When $p_l + x$ is prompted, $c$ is trained to predict $x$'s ground truth label $y$ directly.

%During inference, for any given query $x$, the label generation prompt $p_l$ is prepended to $x$ to elicit $c$ to generate the corresponding label.

%\subsection{Advantages of GARAGE}
%In summary, we present the following highlights of our proposed GARAGE:

%Since any sample of the three datasets is represented in the conditional generation format $(x, y)$, the client-side LM $c$ can select a variety of architectures, including both encoder-decoder and decoder-only transformers to maximize $c$'s conditional generation probability $\text{Pr}(y | x)$.
%F

\subsection{Extending FDKT to One-to-many Scenario with Multiple Clients}
In addition to the one-to-one server-client configuration for the client SLM's own improvement, in this section, we extend FDKT to support the one-to-many scenarios with diverse tasks from multiple clients.
This extension enables individual clients to train multi-task SLMs to handle other clients' tasks simultaneously.

As depicted in the right subpart of Figure~\ref{fig:garage}, following the one-to-one configuration, client $i$ transfers its synthetic data $D_i'$ to the server.
Then, the server LLM $S$ performs in-context data augmentation based on the filtered data $\mathcal{F}(D_i')$ to generate the augmented data $D_i^a$.
To facilitate multi-task training for various clients, the server maintains a multi-task prefix processor which assigns a task-specific prefix for each $D_i^a$.
Depending on clients' tasks, prefixes can be different across different clients and are prepended to the input $x$ in each data pair $(x,y)$ within $D_i^a$.
Following this, the server aggregates all $D_i^a$ to form the final augmented data $D^a$ for multi-task training.
In the final step, the server dispatches both $D^a$ and all prefixes back to clients for fine-tuning local SLMs with language modeling objectives.
During inference, by inserting the appropriate prefixes at the beginning of inputs, the tuned SLM can be utilized for designated tasks.

\section{Experiments}
\label{sec: exp}
To evaluate the effectiveness of the proposed FDKT, we conduct comprehensive experiments, and the details of our experiments are introduced below.

\subsection{Experimental Setups}
\label{sec:exp setup}
\textbf{Datasets}.
Following prior works~\cite{yue2022synthetic, Kurakin2023HarnessingLM}, we conduct our experiments on the Yelp dataset~\cite{Zhang2015CharacterlevelCN} for review rating prediction. 
We sample our experimented data from three domains of the Yelp dataset, including \textit{Shopping}, \textit{Art}, and \textit{Health}.
For each review, we retain its review text and rating. 
%The primary objective is to accurately predict the numerical rating based solely on the review text.
Beyond review classification, we also include the AGNews~\cite{Zhang2015CharacterlevelCN} dataset to predict the news topic.
%\tbc{AGNews....}

\textbf{Data Split}.
In each domain of \textit{Shopping}, \textit{Art}, and \textit{Health},  we filter 5,000 samples and enforce a uniform distribution across all 5 categories to establish balanced datasets. 
For \textit{AGNews}, we sample 5,000 records that are distributed uniformly over 4 topic labels.
We randomly select 1,000 non-overlapped data points for each subset as testing data to report evaluation results.

%\tbc{AGNews....}

\textbf{FDKT details}.
For each domain, we first use generator $G$ to sample 20,000 synthetic samples and apply the filter $\mathcal{F}$ to select 7,000 samples.
We then augment 30,000 examples based on $\mathcal{F}(D')$.
For generator $G$'s privacy budgets, we fix $\epsilon$ = 8 and $\delta$ = 1e-5.

\textbf{Evaluated Models}.
%We introduce the local models and server-side LLM, respectively. 
Our evaluated models include different model architectures for both local SLMs and server-side LLMs.
For local models, we use DP-tuned GPT-2\textsubscript{large} \cite{radford-2019-language} as our generator $G$ to generate synthetic data and use pre-trained T5\textsubscript{large} \cite{2020t5} as the client SLM $c$. 
We follow~\cite{2020t5} to consider the rating prediction as a seq2seq task.
For server-side LLMs $S$, we use Llama-3\textsubscript{8B}~\cite{llama3modelcard} for main experiments.
In addition, we also report FDKT's performance over a wide range of opensource LLMs including Mistral\textsubscript{7B} \cite{Jiang2023Mistral7}, Llama-2\textsubscript{7B}~\cite{touvron2023llama},Qwen\textsubscript{7B} and Qwen\textsubscript{14B}~\cite{qwen}.
% \tbc{For server-side LLMs $S$, we use Mistral-7B \cite{Jiang2023Mistral7}, Llama-2\textsubscript{7B}, Llama-2\textsubscript{13B}~\cite{touvron2023llama}, Vicuna\textsubscript{7B}, Vicuna\textsubscript{13B} \cite{Zheng2023JudgingLW}, Qwen\textsubscript{7B}, Qwen\textsubscript{14B}~\cite{qwen} to explore the impact of different LLMs on our framework.
% }

\textbf{Evaluation Metrics}. 
%\lhr{Two evaluation on generation and logits based are still missing....}
To evaluate SLMs' performance, we perform greedy decoding on the testing data and use regular expressions to extract the generated labels.
All extraction failures are regarded as incorrect predictions.
We report the \textit{Exact Acc} that calculates the exact prediction accuracy for ground truth labels.
In addition, since our Yelp reviews have 5-scale ratings, we aggregate the five rating labels into three sentiment categories: positive, neutral, and negative and report \textit{Rough Acc} as the accuracy for these 3 labels. 
Specifically, ratings of 1-2 stars are classified as negative, 3 stars as neutral, and 4-5 stars as positive.
Throughout our experiments, we report both accuracies in \%.

\textbf{Training details}.
Unless otherwise specified, we optimize \textit{Local FT} for 100 epochs and train SLMs of \textit{Syn FT}, \textit{Gen KT} and FDKT for 20 epochs to report the evaluation results.
For full experimental details, please refer to Appendix~\ref{app: train detail}.

%For the Yelp dataset, each review has a 5-scale rating from 1 to 5 stars.
%We consider two accuracy metrics for the review rating prediction.
%One is the commonly used accuracy metric. The other one is `rough evaluation accuracy'. 
% The first one is the exact accuracy (\textit{Exact Acc}) that calculates the exact prediction accuracy for the 5 labels.
% The second metric, \textit{Rough Acc}, assesses the model's rough accuracy by aggregating the five rating labels into three broad sentiment categories: positive, neutral, and negative. Specifically, ratings of 1-2 stars are classified as negative, 3 stars as neutral, and 4-5 stars as positive.
%We observe that in the Yelp dataset, the review data labeled '1' and '2' means negative while '4' and '5' indicate positive review, and there is no significant difference between them. So we merge '1' and '2' as 'negative', '4' and '5' as 'positive', '3' as 'neutral' to roughly evaluate our method. And the rough evaluation further demonstrates the superiority of our method.

\subsection{Baselines}
We consider the following three baselines to compare our proposed FDKT:

\textbullet  \textbf{Local FT}:
Local FT refers to the local fine-tuning baseline that directly fine-tunes the client SLM $c$ on private data $D$ without any additional data.

\textbullet  \textbf{Syn FT}:
Syn FT denotes synthetic fine-tuning~\cite{yue2022synthetic, Kurakin2023HarnessingLM} that fine-tunes $c$ on the combination of synthetic data $D'$ and the client's private data $D$.

\textbullet  \textbf{Syn FT+$\mathcal{F}$}:
Syn FT+$\mathcal{F}$ applies the data filtering mechanism $\mathcal{F}$ on the synthetic data $D'$ as mentioned in Section~\ref{sec: selection}. Then, the client SLM $c$ is fine-tuned on the combination of filtered data $\mathcal{F}(D')$ and the client's private data $D$.

\textbullet  \textbf{Gen KT}:
Gen KT represents the general knowledge transfer pipeline~\cite{wang2023can, deng2023mutual, ye-2022-zerogen} that leverages LLM $S$'s knowledge on its pre-training data and performs zero-shot data augmentation by only providing necessary descriptions about private data $D$'s tasks' and labels' information.
We use $D^g$ to denote Gen KT's augmented data.
The client SLM $c$ is fine-tuned on the combination of $D^g$ and $D$.

%To ensure fair comparisons between \textit{Gen KT} and our $FDKT$, we set $|D^g|$ = $|D^a|$ = 30,000.

\subsection{Experimental Results}

{
\setlength{\tabcolsep}{3.8pt} % Default is 6pt
\begin{table*}[t]
\centering
%\small
%\fontsize{9pt}{9pt}\selectfont
%\setlength\extrarowheight{1pt}

  \begin{tabular}{l | cc| cc| cc| c}
    \toprule
    %{} & {} & {} & {} & {} & {} & {GARAGE}
    %   \\
       \multirow{2}{*}{\textbf{Method}} & 
       \multicolumn{2}{c|}{\textbf{Arts}} & \multicolumn{2}{c|}{\textbf{Health}} & \multicolumn{2}{c|}{\textbf{Shopping}} & \multicolumn{1}{c}{\textbf{AGNews}}\\
       {}  & {Exact} & {Rough} & {Exact} & {Rough} & {Exact} & {Rough} & {Exact}\\
      \midrule

    %% baseline 1
    Local FT  & 54.66\textsubscript{$\pm$4.57} & 70.22\textsubscript{$\pm$4.99} & 55.82\textsubscript{$\pm$1.93} & 81.30\textsubscript{$\pm$0.39} & 50.08\textsubscript{$\pm$2.21} & 70.30\textsubscript{$\pm$3.28} & 73.57\textsubscript{$\pm$7.62}\\

    %%%%% $D'$ + $D$
    Syn FT & 52.57\textsubscript{$\pm$3.29} & 64.73\textsubscript{$\pm$4.80} & 52.28\textsubscript{$\pm$5.97} & 72.76\textsubscript{$\pm$7.33} & 47.82\textsubscript{$\pm$3.73} & 65.72\textsubscript{$\pm$5.18} & 74.45\textsubscript{$\pm$8.85}\\
    
    %%% $\mathcal{F}(D')  + D$ 
    Syn FT+$\mathcal{F}$  & 55.72\textsubscript{$\pm$3.16} & 72.68\textsubscript{$\pm$2.88} & 55.72\textsubscript{$\pm$3.15} & 75.96\textsubscript{$\pm$3.66} & 50.86\textsubscript{$\pm$3.26} & 67.98\textsubscript{$\pm$4.60} & 76.95\textsubscript{$\pm$3.70}\\
%\cite{deng2023mutual}
    Gen KT   & 60.10\textsubscript{$\pm$0.83} & 79.20\textsubscript{$\pm$2.04} & 54.17\textsubscript{$\pm$3.36} & 82.13\textsubscript{$\pm$0.05} & 53.80\textsubscript{$\pm$2.67} & 74.55\textsubscript{$\pm$2.59} & 86.97\textsubscript{$\pm$2.51}\\
    
    FDKT  & \textbf{62.87\textsubscript{$\pm$2.45}} & \textbf{80.97\textsubscript{$\pm$1.30}} & \textbf{56.43\textsubscript{$\pm$1.53}} & \textbf{82.23\textsubscript{$\pm$0.33}} & \textbf{56.13\textsubscript{$\pm$0.57}} & \textbf{78.43\textsubscript{$\pm$0.45}} & \textbf{87.83\textsubscript{$\pm$1.53}} \\

    \bottomrule
  \end{tabular}
\vspace{-0.05in}
\caption{\label{tab:main_individual}
Evaluation results for the one-to-one scenario where there is one server and one client.
Syn FT+$\mathcal{F}$ refers to fine-tuning on the filtered synthetic data. Exact and Rough denote the exact and rough accuracy, respectively.
Except Local FT, for all other methods, we fine-tune client SLMs on both private data and generated data.
All results are reported in \%.
}

\vspace{-0.15in}
\end{table*}
}
\subsubsection{Evaluation on the One-to-one Scenario}
We explore the effectiveness of FDKT in multiple domains in terms of improvement in individual domains' local SLMs.
Within each domain, we conduct a comparative evaluation of FDKT against \textit{Local FT}, \textit{Syn FT} and \textit{Gen KT} for both exact and rough accuracy.
We generate 20,000 synthetic samples $D'$ and retain 7,000 samples for $\mathcal{F}(D')$.
Then, we exploit server LLM $S$ to augment 30,000 samples as $D^a$.
To ensure fair comparisons, we randomly sample from $D'$ to set $|D'|$=7,000 and we also set $|D^g|$=30,000 for \textit{Gen KT}.
During training, we mix private data with generated data.

The evaluation results are shown in Table~\ref{tab:main_individual}, where we train clients' SLMs over 5 random seeds and report their mean accuracies and sample standard deviation.
The results imply the following findings:

1): \textit{FDKT consistently achieves superior performance across all evaluated domains with less variance.}
Both \textit{Syn FT} and \textit{Gen KT} under-perform \textit{Local FT} occasionally for Health and Shopping domain.
Instead, FDKT always outperforms \textit{Local FT} and other baselines over the 4 domains, achieving the highest results in both exact and rough accuracies.
For example, in the domains of Arts and Shopping, although we train the Local FT for 100 epochs to optimize its performance, FDKT outperforms \textit{Local FT} by 5\% and 7\%  in \textit{Exact Acc} and \textit{Rough Acc}, respectively.
For AGNews, FDKT even gains 14\% improvement over \textit{Local FT}.
The consistent improvements suggest that FDKT is capable of enhancing the task-specific performance of the client SLM $c$.

2): \textit{Synthetic data fail to improve client SLMs' task-specific performance.}
Our results indicate that \textit{Syn FT} and \textit{Syn FT+$\mathcal{F}$} offer only marginal improvements over \textit{Local FT} and sometimes even worsen SLMs' performance.
Moreover, \textit{Syn FT} leads to unstable performance with higher variance.
Such high variance is likely to be caused by the generator $G$'s injected noise.

3): \textit{Our filtering mechanism $\mathcal{F}$ can effectively mitigate synthetic data $D'$'s negative impacts.}
Due to compromised data quality and homogeneous data distribution, \textit{Syn FT} frequently leads to the worst performance even though the SLMs are fine-tuned on $D'$ + $D$.
After adding the filter mechanism $\mathcal{F}$,  \textit{Syn FT+$\mathcal{F}$} fine-tuned on $\mathcal{F}(D')$ + $D$ leads to better accuracies with smaller variance. 
Such improvements emphasize $\mathcal{F}$'s effectiveness in enhancing synthetic data quality, making it a valuable component for our FDKT's pipeline.

%tab:main_individual
% The 2 bar charts in Figure~\ref{fig:domain} show the two accuracy evaluation results of \textit{Local FT}, \textit{Syn FT} and FDKT across four distinct domains: \textit{Shopping}, \textit{Art}, \textit{Health}, and \textit{AGNews}. 
% By comparing two evaluated accuracies on \textit{Local FT} and \textit{Syn FT} with an equivalent volume of fine-tuning data, \textit{Syn FT} exhibits marginally lower performance than \textit{Local FT}.
% This result indicates that quality of synthetic data for causal language modeling may be compromised by its noisy optimization process.

% Regarding the performance of FDKT, it is consistently superior across all evaluated domains.
% The consistent improvements suggest that FDKT is capable of selectively acquiring the necessary knowledge to enhance the task-specific performance of the client LM $c$.
% In addition, we observe that FDKT's advantages are particularly noticeable in scenarios where the private data $D$ is limited.
% For instance, within the \textit{Shopping} domain where $|D| = 20,000$, FDKT achieves an approximate 2\% increase in accuracy. 
% In the \textit{Health} domain, with a smaller dataset of $|D| = 5,000$, the accuracy boost is around 10\%. 
% These substantial gains in accuracies highlight FDKT's potential for significant improvements under data-scarce conditions.

{
\setlength{\tabcolsep}{5pt} % Default is 6pt
\begin{table*}[t]
\centering
%\small
%\fontsize{9pt}{9pt}\selectfont
%\setlength\extrarowheight{1pt}

  \begin{tabular}{l l| cc| cc| cc| c}
    \toprule
    %{} & {} & {} & {} & {} & {} & {GARAGE}
    %   \\
       \multirow{2}{*}{\textbf{Method}} & 
       \multirow{2}{*}{\textbf{FT Data}} &
       \multicolumn{2}{c|}{\textbf{Arts}} & \multicolumn{2}{c|}{\textbf{Health}} & \multicolumn{2}{c|}{\textbf{Shopping}} & \multicolumn{1}{c}{\textbf{AGNews}}\\
       {}  & {}  & {Exact} & {Rough} & {Exact} & {Rough} & {Exact} & {Rough} & {Exact}\\
      \midrule

    %%%%% $D'$ + $D$
    %Syn FT & $D'$ & xx.xx & xx.xx & xx.xx & xx.xx & xx.xx & xx.xx\\
    
    %%% $\mathcal{F}(D')  + D$ 
    %Syn FT+$\mathcal{F}$ & $\mathcal{F}(D')$ & xx.xx & xx.xx & xx.xx & xx.xx & xx.xx & xx.xx\\

    Gen KT & $D^g$ & 32.50 & 53.10 & 39.70 & 59.70 & 31.40 & 47.20 & 65.30\\
    
    FDKT & $D^a$ & \textbf{42.20} & \textbf{62.00} & \textbf{52.40} & \textbf{77.90} & \textbf{44.90} & \textbf{66.60} & \textbf{75.60}\\

    \bottomrule
  \end{tabular}
\vspace{-0.05in}
\caption{\label{tab:main_quality}
Evaluation of the quality of data generated by Gen KT and FDKT with 30,000 augmented data.
FT Data denotes the data used for fine-tuning client SLMs.
All results are reported in \%.
}

\vspace{-0.1in}
\end{table*}
}
\subsubsection{Evaluation on the Quality of Generated Data}
Besides studying whether FDKT is beneficial for client SLMs' performance, in this section, we compared the quality of data generated by \textit{Gen KT} and FDKT.
Instead of fine-tuning the SLMs based on mixed private data and generated data, we fine-tune SLMs based only on the generated data for each method.
To make a fair comparison, we set  $|D^g|$ = $|D^a|$ = 30,000.

Table~\ref{tab:main_quality} displays client SLMs' performance fine-tuned exclusively on generated data.
According to the results in Table~\ref{tab:main_individual} where private data is also trained, our FDKT only yields about 1\textasciitilde2\% accuracy improvement over \textit{Gen KT}.
However, in the absence of $D$, our FDKT can outperform \textit{Gen KT} by more than 10\%. 
This substantial improvement suggests that FDKT's augmented data more closely matches the distribution of private data.

{
\setlength{\tabcolsep}{6pt} % Default is 6pt
\begin{table*}[t]
\centering
%\small
%\fontsize{9pt}{9pt}\selectfont
%\setlength\extrarowheight{1pt}

  \begin{tabular}{l l| cc| cc}
    \toprule
    %{} & {} & {} & {} & {} & {} & {GARAGE}
    %   \\
       \multirow{2}{*}{\textbf{Private Data \#}} & 
       \multirow{2}{*}{\textbf{Augmented Data \#}} &
       \multicolumn{2}{c|}{\textbf{Exact  Acc (\%)}} & \multicolumn{2}{c}{\textbf{Rough  Acc (\%)}}\\
       {}  & {}  & {Local FT} & {FDKT} & {Local FT} & {FDKT} \\
      \midrule

    %%%%% $D'$ + $D$
    %100 & 600 & 39.90 & 40.70 & 57.90 & 55.90 \\
    
    200 & 1,200 & 46.50 & \textbf{50.20} & 72.50 & \textbf{74.20} \\

    500 & 3,000 & 47.50 & \textbf{51.10} & 72.60 & \textbf{76.90} \\
    
    1,000 & 6,000 & 47.00 & \textbf{51.80} & 75.40 & \textbf{77.00} \\

    2,000 & 12,000 & 51.30 & \textbf{58.70} & 77.10 & \textbf{79.40} \\

    5,000 & 30,000 & 55.82 & \textbf{56.43} & 81.30 & \textbf{82.23} \\

    \bottomrule
  \end{tabular}
\vspace{-0.05in}
\caption{\label{tab:private_nums}
Evaluation results with different numbers of private data for the Health domain.
}

\vspace{-0.15in}
\end{table*}
}
\subsubsection{Evaluation on Extreme Data Scarcity}
In this section, we show our FDKT's effectiveness in tackling clients' data scarcity issues. 
We experiment on Yelp's Health domain with $|D|$ = 200, 500, 1,000, 2,000 and 5,000.
For each $D$, we set $|D^a| = 6 \times |D|$  and fine-tune SLMs for the same iterations mentioned in Section~\ref{sec:exp setup}.
For example, when $|D|$ = 1,000, we Fine-tune \textit{Local FT} for 500 epochs and FDKT for 100 epochs.

Table~\ref{tab:private_nums} depicts evaluation results on \textit{Local FT} and FDKT for various $|D|$.
Both exact and rough accuracies verify that FDKT is effective even when the private data is extremely scarce.

%\newpage
\begin{figure*}[t]
\centering
\includegraphics[width=0.98\textwidth]{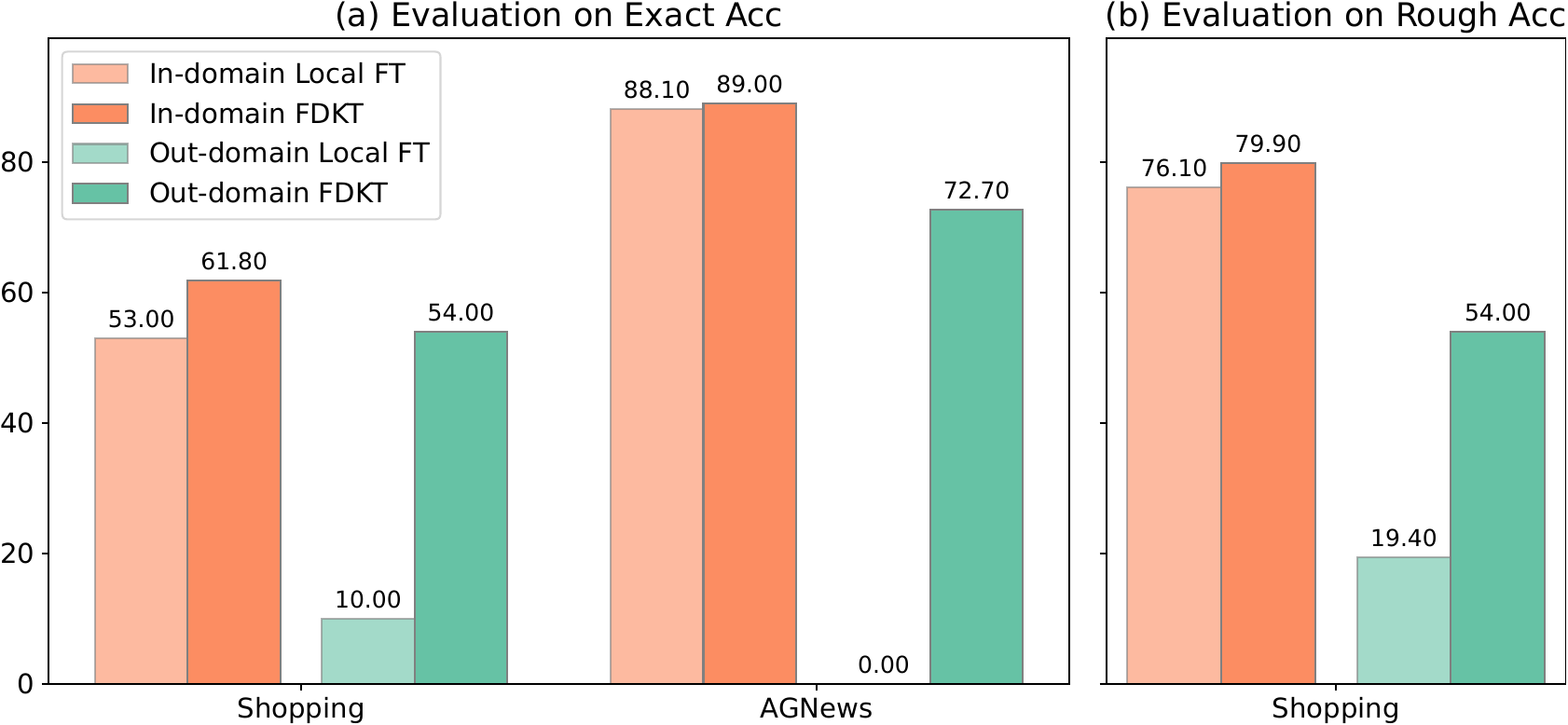}
\vspace{-0.05in}
\caption{
Evaluation of FDKT for the one-to-many scenario.
In-domain Local FT denotes the SLM is fine-tuned and evaluated within the same domain and Out-domain FDKT refers to the SLM fine-tuned on one domain's private data mixed with augmented data $D^a$ and tested on another domain.
}
\label{fig:domain}
\vspace{-0.1in}
\end{figure*}
\subsubsection{Evaluation on the One-to-many Scenario}
Besides the one-to-one configuration, we also study FDKT's effectiveness over multiple clients for multi-task learning. 
For simplicity, we focus on two distinct domains including Shopping and AGNews. 
Following the experimental settings in Section~\ref{sec:exp setup}, each of them serves as a separate client engaged in different tasks.
We merge both domains' 30,000 augmented data samples to obtain $|D^a| = 60,000$ and use their testing data to report in-domain and out-domain results for \textit{Local FT} and FDKT. The term ``in-domain'' indicates that the SLM is fine-tuned and tested on the same domain, while ``out-domain'' refers to testing on the SLM fine-tuned by a different domain.

Figure~\ref{fig:domain} depicts evaluated in-domain and out-domain results for the two clients.
The huge performance gap between \textit{Out-domain Local FT} and \textit{Out-domain FDKT} indicates FDKT's effectiveness in improving clients' SLMs multi-task ability to handle other clients' tasks.
Moreover, after comparing \textit{In-domain Local FT} with \textit{In-domain FDKT}, we observe that FDKT's multi-task learning also improves clients' own task performance.

%%%app:case   app:ablation
\subsubsection{Other Experiments}

\begin{wraptable}{r}{0.5\textwidth}
\centering
%\small
%\fontsize{9pt}{9pt}\selectfont
%\setlength\extrarowheight{1pt}
  \begin{tabular}{l | cc}
    \toprule
    %{} & {} & {} & {} & {} & {} & {GARAGE}
    %   \\
       {LLM Name} & {Exact Acc}  & {Rough Acc}\\
      \midrule

    \multirow{1}{*}{Mistral-7b} 
    & 58.30 & 79.00\\
    \midrule 
    \multirow{1}{*}{Llama2-7b-chat}
    & 55.00 & 74.00\\
    \multirow{1}{*}{Llama3-8b-instruct}
    & 56.13 & 78.43\\
    \midrule 
    \multirow{1}{*}{Qwen-7b-Chat}
    & 54.80 & 76.10\\

    \multirow{1}{*}{Qwen-14b-Chat}
    & 53.60 & 76.10\\

    \bottomrule
  \end{tabular}
\caption{\label{tab:llms}
Evaluation of FDKT's performance over different LLMs within the "Shopping" domain.
}
\vspace{-0.15in}
\end{wraptable}

\textbf{Evaluation over Multiple LLMs}
Beyond different domains, we also extend the evaluation of FDKT's effectiveness to encompass various server-side LLMs.
To maintain consistency in our assessment, we experiment on the \textit{Shopping} domain and set $D^a$ = 30,000 to report accuracies of client SLMs fine-tuned on $D^a$ +$D$.
Our evaluation includes several open-source LLMs of different model sizes and versions, including Mistral~\cite{Jiang2023Mistral7}, Llama-2~\cite{touvron2023llama}, %Vicuna~\cite{zheng2023judging} 
and Qwen v1.5~\cite{qwen}.

The results of this evaluation are summarized in Table~\ref{tab:llms}, where we present \textit{Exact Acc} and \textit{Rough Acc} across these different LLMs.
The results suggest that FDKT integrated with different LLMs consistently surpasses \textit{Local FT} over more than 5\%.
Moreover, FDKT's performance is highly correlated with the server LLM $S$'s capabilities.
FDKT can also benefit from $S$'s improved utility.

\textbf{Case Studies}. We perform case studies including an error analysis about FDKT's $D^a$ and compare a few data samples from $D$,$D'$, $D^a$ and $D^g$.
Detailed results can be found in Appendix~\ref{app:case}.

\textbf{Ablation Studies}. We also evaluate our FDKT's performance with varied privacy budgets and numbers of augmented data. Details can be found in Appendix~\ref{app:ablation}.

% \subsubsection{Ablation Studies}
% To show the effectiveness of FDKT's components, we perform ablation studies with varied numbers of synthetic samples $|D'|$ and augmented samples $|D^a|$.
% We also consider multiple configurations that fine-tune $c$ on synthetic data $D'$ (\textit{Syn}), augmentation $D^a$ (textit{Aug}), both synthetic and augmentation (textit{Syn + Aug}), and the whole FDKT framework (\textit{Syn + Aug + COT}).

% %Table~\ref{tab:ablation} shows the ablation study results on the \textit{Art} domain with $|D| = 10,000$.
% Based on the results, we have the following observations.
% 1): Increasing both $|D'|$ and $|D^a|$ has a general trend of improvement for all configurations.
% 2): \textit{Syn} has the worst performance, which suggests that synthetic samples' quality may be compromised by the injected DP noise and textit{Aug} can improve the data quality with better performance. 
% 3): The combination of $D'$ and $D^a$ is beneficial for $c$ to surpass the \textit{Local FT} baseline.
% 4): \lhr{The distillation step further improves the performance in the majority of cases, indicating that the COT process contributes positively to the model's accuracy.}
% In summary, the ablation study indicates that using a combination of synthetic and augmented data improves model performance, and this effect is further enhanced by distillation.

\section{Conclusion}
In this paper, we explore the federated transfer learning scenarios involving server-side LLMs and client-side SLMs.
Upon identifying the limitations of differentially private synthetic data and general knowledge transfer pipelines, we present the Federated Domain-specific Knowledge Transfer (FDKT) framework.
Instead of directly transferring clients' data to the server which may lead to privacy leakage, we propose to share synthetic data sampled from the differentially private generator $G$ that distributes similarly as the private data.
Then, we propose a data filtering mechanism based on server LLM's data quality evaluation to enhance data quality and discard noisy data compromised by DP.
Finally, the server can perform in-context data augmentation and send back the augmented data for selective knowledge transfer.
Consequently, without any expert annotation, we realize the oriented federated knowledge transfer to improve clients' local SLMs' task-specific performance.
For future work, we aim to expand our framework to incorporate more clients with diverse tasks to train a multi-task SLM collaboratively, potentially increasing the robustness and utility of clients' SLMs.
\newpage

\begin{comment}
\begin{ack}
Use unnumbered first level headings for the acknowledgments. All acknowledgments
go at the end of the paper before the list of references. Moreover, you are required to declare
funding (financial activities supporting the submitted work) and competing interests (related financial activities outside the submitted work).
More information about this disclosure can be found at: \url{https://neurips.cc/Conferences/2023/PaperInformation/FundingDisclosure}.

Do {\bf not} include this section in the anonymized submission, only in the final paper. You can use the \texttt{ack} environment provided in the style file to autmoatically hide this section in the anonymized submission.
\end{ack}
\end{comment}

%%% self added pkgs
\bibliographystyle{plain} % 
\bibliography{references} % This corresponds to 'references.bib'

\begin{thebibliography}{10}

\bibitem{Abadi2016DeepLW}
Mart{\'i}n Abadi, Andy Chu, Ian~J. Goodfellow, H.~B. McMahan, Ilya Mironov, Kunal Talwar, and Li~Zhang.
\newblock Deep learning with differential privacy.
\newblock {\em Proceedings of the 2016 ACM SIGSAC Conference on Computer and Communications Security}, 2016.

\bibitem{llama3modelcard}
AI@Meta.
\newblock Llama 3 model card.
\newblock 2024.

\bibitem{babakniya2023slora}
Sara Babakniya, Ahmed Elkordy, Yahya Ezzeldin, Qingfeng Liu, Kee-Bong Song, MOSTAFA EL-Khamy, and Salman Avestimehr.
\newblock {SL}o{RA}: Federated parameter efficient fine-tuning of language models.
\newblock In {\em International Workshop on Federated Learning in the Age of Foundation Models in Conjunction with NeurIPS 2023}, 2023.

\bibitem{qwen}
Jinze Bai, Shuai Bai, Yunfei Chu, Zeyu Cui, Kai Dang, Xiaodong Deng, Yang Fan, Wenbin Ge, Yu~Han, Fei Huang, Binyuan Hui, Luo Ji, Mei Li, Junyang Lin, Runji Lin, Dayiheng Liu, Gao Liu, Chengqiang Lu, Keming Lu, Jianxin Ma, Rui Men, Xingzhang Ren, Xuancheng Ren, Chuanqi Tan, Sinan Tan, Jianhong Tu, Peng Wang, Shijie Wang, Wei Wang, Shengguang Wu, Benfeng Xu, Jin Xu, An~Yang, Hao Yang, Jian Yang, Shusheng Yang, Yang Yao, Bowen Yu, Hongyi Yuan, Zheng Yuan, Jianwei Zhang, Xingxuan Zhang, Yichang Zhang, Zhenru Zhang, Chang Zhou, Jingren Zhou, Xiaohuan Zhou, and Tianhang Zhu.
\newblock Qwen technical report.
\newblock {\em arXiv preprint arXiv:2309.16609}, 2023.

\bibitem{Bindschaedler-2017-Plausible}
Vincent Bindschaedler, Reza Shokri, and Carl Gunter.
\newblock Plausible deniability for privacy-preserving data synthesis.
\newblock {\em Proceedings of the VLDB Endowment}, 10:481--492, 08 2017.

\bibitem{Brown2020LanguageMA}
Tom~B. Brown, Benjamin Mann, Nick Ryder, Melanie Subbiah, Jared Kaplan, Prafulla Dhariwal, Arvind Neelakantan, Pranav Shyam, Girish Sastry, Amanda Askell, Sandhini Agarwal, Ariel Herbert-Voss, Gretchen Krueger, T.~J. Henighan, Rewon Child, Aditya Ramesh, Daniel~M. Ziegler, Jeff Wu, Clemens Winter, Christopher Hesse, Mark Chen, Eric Sigler, Mateusz Litwin, Scott Gray, Benjamin Chess, Jack Clark, Christopher Berner, Sam McCandlish, Alec Radford, Ilya Sutskever, and Dario Amodei.
\newblock Language models are few-shot learners.
\newblock {\em ArXiv}, abs/2005.14165, 2020.

\bibitem{cai2022autofednlp}
Dongqi Cai, Yaozong Wu, Shangguang Wang, Felix~Xiaozhu Lin, and Mengwei Xu.
\newblock Autofednlp: An efficient fednlp framework.
\newblock {\em arXiv preprint arXiv:2205.10162}, 2022.

\bibitem{Chen2021EvaluatingLL}
Mark Chen, Jerry Tworek, Heewoo Jun, Qiming Yuan, Henrique Ponde, Jared Kaplan, Harrison Edwards, Yura Burda, Nicholas Joseph, Greg Brockman, Alex Ray, Raul Puri, Gretchen Krueger, Michael Petrov, Heidy Khlaaf, Girish Sastry, Pamela Mishkin, Brooke Chan, Scott Gray, Nick Ryder, Mikhail Pavlov, Alethea Power, Lukasz Kaiser, Mohammad Bavarian, Clemens Winter, Philippe Tillet, Felipe~Petroski Such, David~W. Cummings, Matthias Plappert, Fotios Chantzis, Elizabeth Barnes, Ariel Herbert-Voss, William~H. Guss, Alex Nichol, Igor Babuschkin, S.~Arun Balaji, Shantanu Jain, Andrew Carr, Jan Leike, Joshua Achiam, Vedant Misra, Evan Morikawa, Alec Radford, Matthew~M. Knight, Miles Brundage, Mira Murati, Katie Mayer, Peter Welinder, Bob McGrew, Dario Amodei, Sam McCandlish, Ilya Sutskever, and Wojciech Zaremba.
\newblock Evaluating large language models trained on code.
\newblock {\em ArXiv}, abs/2107.03374, 2021.

\bibitem{2022flant5}
Hyung~Won Chung, Le~Hou, S.~Longpre, Barret Zoph, Yi~Tay, William Fedus, Eric Li, Xuezhi Wang, Mostafa Dehghani, Siddhartha Brahma, Albert Webson, Shixiang~Shane Gu, Zhuyun Dai, Mirac Suzgun, Xinyun Chen, Aakanksha Chowdhery, Dasha Valter, Sharan Narang, Gaurav Mishra, Adams~Wei Yu, Vincent Zhao, Yanping Huang, Andrew~M. Dai, Hongkun Yu, Slav Petrov, Ed~Huai hsin Chi, Jeff Dean, Jacob Devlin, Adam Roberts, Denny Zhou, Quoc~V. Le, and Jason Wei.
\newblock Scaling instruction-finetuned language models.
\newblock {\em ArXiv}, abs/2210.11416, 2022.

\bibitem{collins2023profit}
Liam Collins, Shanshan Wu, Sewoong Oh, and Khe~Chai Sim.
\newblock Profit: Benchmarking personalization and robustness trade-off in federated prompt tuning.
\newblock In {\em International Workshop on Federated Learning in the Age of Foundation Models in Conjunction with NeurIPS 2023}, 2023.

\bibitem{deng2023mutual}
Yongheng Deng, Ziqing Qiao, Ju~Ren, Yang Liu, and Yaoxue Zhang.
\newblock Mutual enhancement of large and small language models with cross-silo knowledge transfer.
\newblock {\em arXiv preprint arXiv:2312.05842}, 2023.

\bibitem{duan2023flocks}
Haonan Duan, Adam Dziedzic, Nicolas Papernot, and Franziska Boenisch.
\newblock Flocks of stochastic parrots: Differentially private prompt learning for large language models.
\newblock {\em arXiv preprint arXiv:2305.15594}, 2023.

\bibitem{Dwork-08-DP}
C.~{Dwork} and A.~{Roth}.
\newblock The algorithmic foundations of differential privacy.
\newblock In {\em The Algorithmic Foundations of Differential Privacy}, pages 19--20, 2014.

\bibitem{fan2023fate}
Tao Fan, Yan Kang, Guoqiang Ma, Weijing Chen, Wenbin Wei, Lixin Fan, and Qiang Yang.
\newblock Fate-llm: A industrial grade federated learning framework for large language models.
\newblock {\em arXiv preprint arXiv:2310.10049}, 2023.

\bibitem{Feyisetan-20-Privacy}
Oluwaseyi Feyisetan, Borja Balle, Thomas Drake, and Tom Diethe.
\newblock Privacy- and utility-preserving textual analysis via calibrated multivariate perturbations.
\newblock In {\em Proceedings of the 13th International Conference on Web Search and Data Mining}, WSDM '20, page 178–186, New York, NY, USA, 2020. Association for Computing Machinery.

\bibitem{flemings2024differentially}
James Flemings and Murali Annavaram.
\newblock Differentially private knowledge distillation via synthetic text generation.
\newblock {\em arXiv preprint arXiv:2403.00932}, 2024.

\bibitem{fu2023gptscore}
Jinlan Fu, See-Kiong Ng, Zhengbao Jiang, and Pengfei Liu.
\newblock Gptscore: Evaluate as you desire.
\newblock {\em arXiv preprint arXiv:2302.04166}, 2023.

\bibitem{Greshake-2023-not}
Kai Greshake, Sahar Abdelnabi, Shailesh Mishra, Christoph Endres, Thorsten Holz, and Mario Fritz.
\newblock Not what you've signed up for: Compromising real-world llm-integrated applications with indirect prompt injection.
\newblock In {\em Proceedings of the 16th ACM Workshop on Artificial Intelligence and Security}, AISec '23, page 79–90, New York, NY, USA, 2023. Association for Computing Machinery.

\bibitem{Hartigan-clustering}
John~A. Hartigan.
\newblock {\em Clustering Algorithms}.
\newblock John Wiley \& Sons, Inc., USA, 99th edition, 1975.

\bibitem{he2021towards}
Junxian He, Chunting Zhou, Xuezhe Ma, Taylor Berg-Kirkpatrick, and Graham Neubig.
\newblock Towards a unified view of parameter-efficient transfer learning.
\newblock {\em arXiv preprint arXiv:2110.04366}, 2021.

\bibitem{hong2024dpopt}
Junyuan Hong, Jiachen~T. Wang, Chenhui Zhang, Zhangheng LI, Bo~Li, and Zhangyang Wang.
\newblock {DP}-{OPT}: Make large language model your differentially-private prompt engineer.
\newblock In {\em The Twelfth International Conference on Learning Representations}, 2024.

\bibitem{houlsby2019parameter}
Neil Houlsby, Andrei Giurgiu, Stanislaw Jastrzebski, Bruna Morrone, Quentin De~Laroussilhe, Andrea Gesmundo, Mona Attariyan, and Sylvain Gelly.
\newblock Parameter-efficient transfer learning for nlp.
\newblock In {\em International Conference on Machine Learning}, pages 2790--2799. PMLR, 2019.

\bibitem{hu2021lora}
Edward~J Hu, Yelong Shen, Phillip Wallis, Zeyuan Allen-Zhu, Yuanzhi Li, Shean Wang, Lu~Wang, and Weizhu Chen.
\newblock Lora: Low-rank adaptation of large language models.
\newblock {\em arXiv preprint arXiv:2106.09685}, 2021.

\bibitem{pmlr-v162-huang22a}
Wenlong Huang, Pieter Abbeel, Deepak Pathak, and Igor Mordatch.
\newblock Language models as zero-shot planners: Extracting actionable knowledge for embodied agents.
\newblock In Kamalika Chaudhuri, Stefanie Jegelka, Le~Song, Csaba Szepesvari, Gang Niu, and Sivan Sabato, editors, {\em Proceedings of the 39th International Conference on Machine Learning}, volume 162 of {\em Proceedings of Machine Learning Research}, pages 9118--9147. PMLR, 17--23 Jul 2022.

\bibitem{huang2023privacy}
Yangsibo Huang, Samyak Gupta, Zexuan Zhong, Kai Li, and Danqi Chen.
\newblock Privacy implications of retrieval-based language models.
\newblock {\em arXiv preprint arXiv:2305.14888}, 2023.

\bibitem{Igamberdiev-2023-DP-BART}
Timour Igamberdiev and Ivan Habernal.
\newblock Dp-bart for privatized text rewriting under local differential privacy.
\newblock In {\em Findings of the Association for Computational Linguistics: ACL 2023}, page (to appear), Toronto, Canada, 2023. Association for Computational Linguistics.

\bibitem{Jiang2023Mistral7}
Albert~Qiaochu Jiang, Alexandre Sablayrolles, Arthur Mensch, Chris Bamford, Devendra~Singh Chaplot, Diego de~Las~Casas, Florian Bressand, Gianna Lengyel, Guillaume Lample, Lucile Saulnier, L'elio~Renard Lavaud, Marie-Anne Lachaux, Pierre Stock, Teven~Le Scao, Thibaut Lavril, Thomas Wang, Timoth{\'e}e Lacroix, and William~El Sayed.
\newblock Mistral 7b.
\newblock {\em ArXiv}, abs/2310.06825, 2023.

\bibitem{kang2023grounding}
Yan Kang, Tao Fan, Hanlin Gu, Lixin Fan, and Qiang Yang.
\newblock Grounding foundation models through federated transfer learning: A general framework.
\newblock {\em arXiv preprint arXiv:2311.17431}, 2023.

\bibitem{kasiviswanathan-2011-localdp}
Shiva~Prasad Kasiviswanathan, Homin~K Lee, Kobbi Nissim, Sofya Raskhodnikova, and Adam Smith.
\newblock What can we learn privately?
\newblock {\em SIAM Journal on Computing}, 40(3):793--826, 2011.

\bibitem{Kojima2022LargeLM}
Takeshi Kojima, Shixiang~(Shane) Gu, Machel Reid, Yutaka Matsuo, and Yusuke Iwasawa.
\newblock Large language models are zero-shot reasoners.
\newblock In {\em Advances in Neural Information Processing Systems}, volume~35, pages 22199--22213, 2022.

\bibitem{konevcny2016federated2}
Jakub Konecn{\'{y}}, H.~Brendan McMahan, Daniel Ramage, and Peter Richt{\'{a}}rik.
\newblock Federated optimization: Distributed machine learning for on-device intelligence.
\newblock {\em arXiv preprint arXiv:1610.02527}, 2016.

\bibitem{konevcny2016federated}
Jakub Konečný, H.~Brendan McMahan, Felix~X. Yu, Peter Richtarik, Ananda~Theertha Suresh, and Dave Bacon.
\newblock Federated learning: Strategies for improving communication efficiency.
\newblock In {\em NIPS Workshop on Private Multi-Party Machine Learning}, 2016.

\bibitem{Kurakin2023HarnessingLM}
Alexey Kurakin, Natalia Ponomareva, Umar Syed, Liam MacDermed, and A.~Terzis.
\newblock Harnessing large-language models to generate private synthetic text.
\newblock {\em ArXiv}, abs/2306.01684, 2023.

\bibitem{lester2021power}
Brian Lester, Rami Al-Rfou, and Noah Constant.
\newblock The power of scale for parameter-efficient prompt tuning.
\newblock {\em arXiv preprint arXiv:2104.08691}, 2021.

\bibitem{li-etal-2023-multi-step}
Haoran Li, Dadi Guo, Wei Fan, Mingshi Xu, Jie Huang, Fanpu Meng, and Yangqiu Song.
\newblock Multi-step jailbreaking privacy attacks on {C}hat{GPT}.
\newblock In Houda Bouamor, Juan Pino, and Kalika Bali, editors, {\em Findings of the Association for Computational Linguistics: EMNLP 2023}, pages 4138--4153, Singapore, December 2023. Association for Computational Linguistics.

\bibitem{li2023p}
Haoran Li, Dadi Guo, Donghao Li, Wei Fan, Qi~Hu, Xin Liu, Chunkit Chan, Duanyi Yao, and Yangqiu Song.
\newblock P-bench: A multi-level privacy evaluation benchmark for language models.
\newblock {\em arXiv preprint arXiv:2311.04044}, 2023.

\bibitem{li2021prefix}
Xiang~Lisa Li and Percy Liang.
\newblock Prefix-tuning: Optimizing continuous prompts for generation.
\newblock {\em arXiv preprint arXiv:2101.00190}, 2021.

\bibitem{li2022large}
Xuechen Li, Florian Tramer, Percy Liang, and Tatsunori Hashimoto.
\newblock Large language models can be strong differentially private learners.
\newblock In {\em International Conference on Learning Representations}, 2022.

\bibitem{li2023privacy}
Yansong Li, Zhixing Tan, and Yang Liu.
\newblock Privacy-preserving prompt tuning for large language model services.
\newblock {\em arXiv preprint arXiv:2305.06212}, 2023.

\bibitem{li2024leveraging}
Zhen Li, Xiaohan Xu, Tao Shen, Can Xu, Jia-Chen Gu, and Chongyang Tao.
\newblock Leveraging large language models for nlg evaluation: A survey, 2024.

\bibitem{liang2023taskmatrix}
Yaobo Liang, Chenfei Wu, Ting Song, Wenshan Wu, Yan Xia, Yu~Liu, Yang Ou, Shuai Lu, Lei Ji, Shaoguang Mao, et~al.
\newblock Taskmatrix. ai: Completing tasks by connecting foundation models with millions of apis.
\newblock {\em arXiv preprint arXiv:2303.16434}, 2023.

\bibitem{mattern-2022-differentially}
Justus Mattern, Zhijing Jin, Benjamin Weggenmann, Bernhard Schoelkopf, and Mrinmaya Sachan.
\newblock Differentially private language models for secure data sharing.
\newblock In {\em Proceedings of EMNLP 2022}, pages 4860--4873, Abu Dhabi, United Arab Emirates, December 2022. Association for Computational Linguistics.

\bibitem{mcmahan2017communication}
Brendan McMahan, Eider Moore, Daniel Ramage, Seth Hampson, and Blaise~Aguera y~Arcas.
\newblock Communication-efficient learning of deep networks from decentralized data.
\newblock In {\em Proceedings of the AISTATS}, pages 1273--1282, 2017.

\bibitem{OpenAI2023GPT4TR}
OpenAI.
\newblock Gpt-4 technical report.
\newblock {\em ArXiv}, abs/2303.08774, 2023.

\bibitem{ouyang2022training}
Long Ouyang, Jeffrey Wu, Xu~Jiang, Diogo Almeida, Carroll Wainwright, Pamela Mishkin, Chong Zhang, Sandhini Agarwal, Katarina Slama, Alex Gray, John Schulman, Jacob Hilton, Fraser Kelton, Luke Miller, Maddie Simens, Amanda Askell, Peter Welinder, Paul Christiano, Jan Leike, and Ryan Lowe.
\newblock Training language models to follow instructions with human feedback.
\newblock In Alice~H. Oh, Alekh Agarwal, Danielle Belgrave, and Kyunghyun Cho, editors, {\em Advances in Neural Information Processing Systems}, 2022.

\bibitem{Ozdayi2023}
Mustafa Ozdayi, Charith Peris, Jack G.~M. FitzGerald, Christophe Dupuy, Jimit Majmudar, Haidar Khan, Rahil Parikh, and Rahul Gupta.
\newblock Controlling the extraction of memorized data from large language models via prompt-tuning.
\newblock In {\em ACL 2023}, 2023.

\bibitem{Park-2023-Generative}
Joon~Sung Park, Joseph O'Brien, Carrie~Jun Cai, Meredith~Ringel Morris, Percy Liang, and Michael~S. Bernstein.
\newblock Generative agents: Interactive simulacra of human behavior.
\newblock In {\em Proceedings of the 36th Annual ACM Symposium on User Interface Software and Technology}, UIST '23, New York, NY, USA, 2023. Association for Computing Machinery.

\bibitem{qu2021natural}
Chen Qu, Weize Kong, Liu Yang, Mingyang Zhang, Michael Bendersky, and Marc Najork.
\newblock Natural language understanding with privacy-preserving bert.
\newblock In {\em Proceedings of the 30th ACM International Conference on Information \& Knowledge Management}, pages 1488--1497, 2021.

\bibitem{radford-2019-language}
Alec Radford, Jeff Wu, Rewon Child, David Luan, Dario Amodei, and Ilya Sutskever.
\newblock Language models are unsupervised multitask learners.
\newblock {\em ArXiv}, 2019.

\bibitem{2020t5}
Colin Raffel, Noam Shazeer, Adam Roberts, Katherine Lee, Sharan Narang, Michael Matena, Yanqi Zhou, Wei Li, and Peter~J. Liu.
\newblock Exploring the limits of transfer learning with a unified text-to-text transformer.
\newblock {\em Journal of Machine Learning Research}, 21(140):1--67, 2020.

\bibitem{reimers-2019-sentence-bert}
Nils Reimers and Iryna Gurevych.
\newblock Sentence-bert: Sentence embeddings using siamese bert-networks.
\newblock In {\em Proceedings of the 2019 Conference on Empirical Methods in Natural Language Processing}. Association for Computational Linguistics, 11 2019.

\bibitem{sanh2022multitask}
Victor Sanh, Albert Webson, Colin Raffel, Stephen Bach, Lintang Sutawika, Zaid Alyafeai, Antoine Chaffin, Arnaud Stiegler, Arun Raja, Manan Dey, M~Saiful Bari, Canwen Xu, Urmish Thakker, Shanya~Sharma Sharma, Eliza Szczechla, Taewoon Kim, Gunjan Chhablani, Nihal Nayak, Debajyoti Datta, Jonathan Chang, Mike Tian-Jian Jiang, Han Wang, Matteo Manica, Sheng Shen, Zheng~Xin Yong, Harshit Pandey, Rachel Bawden, Thomas Wang, Trishala Neeraj, Jos Rozen, Abheesht Sharma, Andrea Santilli, Thibault Fevry, Jason~Alan Fries, Ryan Teehan, Teven~Le Scao, Stella Biderman, Leo Gao, Thomas Wolf, and Alexander~M Rush.
\newblock Multitask prompted training enables zero-shot task generalization.
\newblock In {\em International Conference on Learning Representations}, 2022.

\bibitem{Shen2023HuggingGPTSA}
Yongliang Shen, Kaitao Song, Xu~Tan, Dong~Sheng Li, Weiming Lu, and Yue~Ting Zhuang.
\newblock Hugginggpt: Solving ai tasks with chatgpt and its friends in hugging face.
\newblock {\em ArXiv}, abs/2303.17580, 2023.

\bibitem{shi-etal-2022-just}
Weiyan Shi, Ryan Shea, Si~Chen, Chiyuan Zhang, Ruoxi Jia, and Zhou Yu.
\newblock Just fine-tune twice: Selective differential privacy for large language models.
\newblock In {\em Proceedings of the 2022 Conference on Empirical Methods in Natural Language Processing}, pages 6327--6340, Abu Dhabi, United Arab Emirates, December 2022. Association for Computational Linguistics.

\bibitem{Significant_Gravitas_AutoGPT}
{Significant Gravitas}.
\newblock {AutoGPT}, 2023.

\bibitem{sun2023principledriven}
Zhiqing Sun, Yikang Shen, Qinhong Zhou, Hongxin Zhang, Zhenfang Chen, David~Daniel Cox, Yiming Yang, and Chuang Gan.
\newblock Principle-driven self-alignment of language models from scratch with minimal human supervision.
\newblock In {\em Thirty-seventh Conference on Neural Information Processing Systems}, 2023.

\bibitem{alpaca}
Rohan Taori, Ishaan Gulrajani, Tianyi Zhang, Yann Dubois, Xuechen Li, Carlos Guestrin, Percy Liang, and Tatsunori~B. Hashimoto.
\newblock Stanford alpaca: An instruction-following llama model.
\newblock \url{https://github.com/tatsu-lab/stanford_alpaca}, 2023.

\bibitem{touvron2023llama}
Hugo Touvron, Louis Martin, Kevin Stone, Peter Albert, Amjad Almahairi, Yasmine Babaei, Nikolay Bashlykov, Soumya Batra, Prajjwal Bhargava, Shruti Bhosale, Dan Bikel, Lukas Blecher, Cristian~Canton Ferrer, Moya Chen, Guillem Cucurull, David Esiobu, Jude Fernandes, Jeremy Fu, Wenyin Fu, Brian Fuller, Cynthia Gao, Vedanuj Goswami, Naman Goyal, Anthony Hartshorn, Saghar Hosseini, Rui Hou, Hakan Inan, Marcin Kardas, Viktor Kerkez, Madian Khabsa, Isabel Kloumann, Artem Korenev, Punit~Singh Koura, Marie-Anne Lachaux, Thibaut Lavril, Jenya Lee, Diana Liskovich, Yinghai Lu, Yuning Mao, Xavier Martinet, Todor Mihaylov, Pushkar Mishra, Igor Molybog, Yixin Nie, Andrew Poulton, Jeremy Reizenstein, Rashi Rungta, Kalyan Saladi, Alan Schelten, Ruan Silva, Eric~Michael Smith, Ranjan Subramanian, Xiaoqing~Ellen Tan, Binh Tang, Ross Taylor, Adina Williams, Jian~Xiang Kuan, Puxin Xu, Zheng Yan, Iliyan Zarov, Yuchen Zhang, Angela Fan, Melanie Kambadur, Sharan Narang, Aurelien Rodriguez, Robert Stojnic, Sergey Edunov, and Thomas
  Scialom.
\newblock Llama 2: Open foundation and fine-tuned chat models, 2023.

\bibitem{wang2023can}
Boxin Wang, Yibo~Jacky Zhang, Yuan Cao, Bo~Li, H~Brendan McMahan, Sewoong Oh, Zheng Xu, and Manzil Zaheer.
\newblock Can public large language models help private cross-device federated learning?
\newblock {\em arXiv preprint arXiv:2305.12132}, 2023.

\bibitem{wang-etal-2023-self-instruct}
Yizhong Wang, Yeganeh Kordi, Swaroop Mishra, Alisa Liu, Noah~A. Smith, Daniel Khashabi, and Hannaneh Hajishirzi.
\newblock Self-instruct: Aligning language models with self-generated instructions.
\newblock In Anna Rogers, Jordan Boyd-Graber, and Naoaki Okazaki, editors, {\em Proceedings of the 61st Annual Meeting of the Association for Computational Linguistics (Volume 1: Long Papers)}, pages 13484--13508, Toronto, Canada, July 2023. Association for Computational Linguistics.

\bibitem{Wei2022ChainOT}
Jason Wei, Xuezhi Wang, Dale Schuurmans, Maarten Bosma, brian ichter, Fei Xia, Ed~H. Chi, Quoc~V Le, and Denny Zhou.
\newblock Chain of thought prompting elicits reasoning in large language models.
\newblock In Alice~H. Oh, Alekh Agarwal, Danielle Belgrave, and Kyunghyun Cho, editors, {\em Advances in Neural Information Processing Systems}, 2022.

\bibitem{west-etal-2022-symbolic}
Peter West, Chandra Bhagavatula, Jack Hessel, Jena Hwang, Liwei Jiang, Ronan Le~Bras, Ximing Lu, Sean Welleck, and Yejin Choi.
\newblock Symbolic knowledge distillation: from general language models to commonsense models.
\newblock In Marine Carpuat, Marie-Catherine de~Marneffe, and Ivan~Vladimir Meza~Ruiz, editors, {\em Proceedings of the 2022 Conference of the North American Chapter of the Association for Computational Linguistics: Human Language Technologies}, pages 4602--4625, Seattle, United States, July 2022. Association for Computational Linguistics.

\bibitem{Williams-1989-teacher}
Ronald~J. Williams and David Zipser.
\newblock A learning algorithm for continually running fully recurrent neural networks.
\newblock {\em Neural Computation}, 1(2):270--280, 1989.

\bibitem{yang2023harnessing}
Jingfeng Yang, Hongye Jin, Ruixiang Tang, Xiaotian Han, Qizhang Feng, Haoming Jiang, Bing Yin, and Xia Hu.
\newblock Harnessing the power of llms in practice: A survey on chatgpt and beyond.
\newblock {\em arXiv preprint arXiv:2304.13712}, 2023.

\bibitem{YangLCT19}
Qiang Yang, Yang Liu, Tianjian Chen, and Yongxin Tong.
\newblock Federated machine learning: Concept and applications.
\newblock {\em {ACM} {TIST}}, 10(2):12:1--12:19, 2019.

\bibitem{ye-2022-zerogen}
Jiacheng Ye, Jiahui Gao, Qintong Li, Hang Xu, Jiangtao Feng, Zhiyong Wu, Tao Yu, and Lingpeng Kong.
\newblock {Z}ero{G}en: Efficient zero-shot learning via dataset generation.
\newblock In Yoav Goldberg, Zornitsa Kozareva, and Yue Zhang, editors, {\em Proceedings of the 2022 Conference on Empirical Methods in Natural Language Processing}, pages 11653--11669, Abu Dhabi, United Arab Emirates, December 2022. Association for Computational Linguistics.

\bibitem{ye2024openfedllm}
Rui Ye, Wenhao Wang, Jingyi Chai, Dihan Li, Zexi Li, Yinda Xu, Yaxin Du, Yanfeng Wang, and Siheng Chen.
\newblock Openfedllm: Training large language models on decentralized private data via federated learning.
\newblock {\em arXiv preprint arXiv:2402.06954}, 2024.

\bibitem{opacus}
Ashkan Yousefpour, Igor Shilov, Alexandre Sablayrolles, Davide Testuggine, Karthik Prasad, Mani Malek, John Nguyen, Sayan Ghosh, Akash Bharadwaj, Jessica Zhao, Graham Cormode, and Ilya Mironov.
\newblock Opacus: {U}ser-friendly differential privacy library in {PyTorch}.
\newblock {\em arXiv preprint arXiv:2109.12298}, 2021.

\bibitem{yu2023selective}
Da~Yu, Sivakanth Gopi, Janardhan Kulkarni, Zinan Lin, Saurabh Naik, Tomasz~Lukasz Religa, Jian Yin, and Huishuai Zhang.
\newblock Selective pre-training for private fine-tuning.
\newblock {\em arXiv preprint arXiv:2305.13865}, 2023.

\bibitem{yu2022differentially}
Da~Yu, Saurabh Naik, Arturs Backurs, Sivakanth Gopi, Huseyin~A Inan, Gautam Kamath, Janardhan Kulkarni, Yin~Tat Lee, Andre Manoel, Lukas Wutschitz, Sergey Yekhanin, and Huishuai Zhang.
\newblock Differentially private fine-tuning of language models.
\newblock In {\em International Conference on Learning Representations}, 2022.

\bibitem{yuan-etal-2023-distilling}
Siyu Yuan, Jiangjie Chen, Ziquan Fu, Xuyang Ge, Soham Shah, Charles Jankowski, Yanghua Xiao, and Deqing Yang.
\newblock Distilling script knowledge from large language models for constrained language planning.
\newblock In Anna Rogers, Jordan Boyd-Graber, and Naoaki Okazaki, editors, {\em Proceedings of the 61st Annual Meeting of the Association for Computational Linguistics (Volume 1: Long Papers)}, pages 4303--4325, Toronto, Canada, July 2023. Association for Computational Linguistics.

\bibitem{yue2022synthetic}
Xiang Yue, Huseyin~A Inan, Xuechen Li, Girish Kumar, Julia McAnallen, Huan Sun, David Levitan, and Robert Sim.
\newblock Synthetic text generation with differential privacy: A simple and practical recipe.
\newblock In {\em Proceedings of ACL 2023}, 2022.

\bibitem{Zhang2015CharacterlevelCN}
Xiang Zhang, Junbo~Jake Zhao, and Yann LeCun.
\newblock Character-level convolutional networks for text classification.
\newblock In {\em NIPS}, 2015.

\bibitem{zhang2022federated}
Zhuo Zhang, Yuanhang Yang, Yong Dai, Lizhen Qu, and Zenglin Xu.
\newblock When federated learning meets pre-trained language models' parameter-efficient tuning methods.
\newblock {\em arXiv preprint arXiv:2212.10025}, 2022.

\bibitem{zhao2022reduce}
Haodong Zhao, Wei Du, Fangqi Li, Peixuan Li, and Gongshen Liu.
\newblock Reduce communication costs and preserve privacy: Prompt tuning method in federated learning.
\newblock {\em arXiv preprint arXiv:2208.12268}, 2022.

\bibitem{zheng2023judging}
Lianmin Zheng, Wei-Lin Chiang, Ying Sheng, Siyuan Zhuang, Zhanghao Wu, Yonghao Zhuang, Zi~Lin, Zhuohan Li, Dacheng Li, Eric Xing, Hao Zhang, Joseph~E. Gonzalez, and Ion Stoica.
\newblock Judging {LLM}-as-a-judge with {MT}-bench and chatbot arena.
\newblock In {\em Thirty-seventh Conference on Neural Information Processing Systems Datasets and Benchmarks Track}, 2023.

\bibitem{zhou2023comprehensive}
Ce~Zhou, Qian Li, Chen Li, Jun Yu, Yixin Liu, Guangjing Wang, Kai Zhang, Cheng Ji, Qiben Yan, Lifang He, et~al.
\newblock A comprehensive survey on pretrained foundation models: A history from bert to chatgpt.
\newblock {\em arXiv preprint arXiv:2302.09419}, 2023.

\bibitem{zhou2023leasttomost}
Denny Zhou, Nathanael Sch{\"a}rli, Le~Hou, Jason Wei, Nathan Scales, Xuezhi Wang, Dale Schuurmans, Claire Cui, Olivier Bousquet, Quoc~V Le, and Ed~H. Chi.
\newblock Least-to-most prompting enables complex reasoning in large language models.
\newblock In {\em The Eleventh International Conference on Learning Representations}, 2023.

\end{thebibliography}

%%%%%%%%%%%%%%%%%%%%%%%%%%%%%%%%%%%%%%%%%%%%%%%%%%%%%%%%%%%%
\appendix
\clearpage

% \section*{Limitations}
% Our limitations can be discussed in two folds.

% \textbf{Privacy}. 
% Although our FDKT transmits synthetic and differentially private data to the server to protect the privacy of the client's sensitive data, the label information is inevitably exposed to the server for data augmentation.
% In addition, the server may potentially infer the client's rough data distribution according to the shared synthetic data.

% \textbf{Utility}. 
% For our FDKT, the client's performance gain relies heavily on the capability of the server LLM.
% Though prompt engineering can enhance the quality of in-context data augmentation, various LLMs respond differently to distinct prompt patterns.
% Hence, optimizing FDKT's performance presents significant challenges.

% \section*{Broader Impacts}
% \label{impact}
% Our proposed FDKT enables a selective and generative federated knowledge transfer pipeline between SLMs and LLMs.
% Compared with existing works, our selective knowledge transfer leads to consistent and significant performance improvements with strict privacy guarantees.
% Moreover, our FDKT can be flexibly adapted to any LMs' architecture and requires only API-level access to the LLMs.

% Based on the above advantages, we expect that our FDKT will have a positive impact on federated learning communities by improving the privacy-utility trade-off for sensitive domains.

\section{More on Training Details}
\label{app: train detail}

\subsection{Hyper-parameters}
%Throughout our experiments, 

\textbf{Synthetic data generator $G$}.
To train the generator $G$ with private data $D$, we use a DP-AdamW optimizer based on the modified Opacus package~\cite{opacus, li2023p} with lr = 1e-5.
We follow~\cite{li2022large} to freeze the token embeddings during the training process.
We set the virtual batch size to 64, the actual batch size equal to 4 and the epoch to 100. 
For DP settings, we set the target $\delta$ = 1e-5, $\epsilon$ = 8 and max\_grad\_norm = 1.
To sample from $G$, we use sampling-based decoding with top\_k = 50, top\_p = 90 and temperature = 1.

\textbf{Client-side SLM $c$}.
To train the encoder-decoder client SLM $c$, we use the AdamW optimizer with lr = 1e-4 and warm\_up\_step = 40.  
We set the virtual batch size to 64, the actual batch size equal to 4 and the epoch to 20. 
During inference, we use greedy decoding to decode the labels given the input texts.
%For random seeds used for reporting Table~\ref{tab:main_individual}, we train clients' SLMs for 5 times with random\_seed = 40, 41, 42, 43, 44, respectively.

\textbf{Server-side LLM $S$}.
To perform data augmentation, we use sampling-based batch decoding.
We set the batch size to 8, temperature=0.6,
and top\_p=0.9.

\subsection{Other Details}

\textbf{Computational Resources}. 
During our experiment, we use 2 Nvidia A100 80GB graphic cards to run our codes and it takes around 30 days of GPU hours to complete all experiments.

\textbf{Full Prompt Templates}. 
Table~\ref{tab:prompt} lists prompt template examples with few-shot demonstrations for data augmentation and data filtering.
%%% LLM prompt for 1) distillation and 2) 

\textbf{Dataset Licenses}.
We use the Yelp dataset under Apache License, Version 2.0, and the AGNews data under Custom (non-commercial) license.

\textbf{Data filtering}.
We use the K-means algorithm for text clustering and choose all-mpnet-base-v2\footnote{\url{https://huggingface.co/sentence-transformers/all-mpnet-base-v2}} as the text embedding model. 
We decide the number of cluster centers so that each cluster has an average of 20 text data points. Then using the prompt in the second row of Table \ref{tab:prompt}, we ask the LLM to remove semantically redundant or ambiguous data, and leave behind high-quality, representative data. Finally, we use regular expressions to retrieve the text indexes selected by the LLM.
%\input{app-individual_domain}
%\section{More Experimental Results}

%\subsection{Experimental Results on Constrained Decoding}

\section{Case Studies}
\label{app:case}

\begin{wraptable}{r}{0.5\textwidth}
    \centering
    \begin{tabular}{ccccccccc}
    \toprule
         \multirow{2}{*}{} & {} & {} & {} & \multicolumn{5}{c}{Prediction}  \\
         \cline{5-9}
         \addlinespace[0.3em]
         {} & {} & {} & {} & 1 & 2 & 3 & 4 & 5 \\
         \hline 
         \addlinespace[0.3em]
         \multicolumn{3}{c}{\multirow{5}{*}{\begin{tabular}{@{}c@{}}Ground \\ Truth\end{tabular}}} & 1 & 844 & 147 & 5 & 1 & 3 \\
         {} & {} & {} & 2 & 384 & 574 & 32 & 7 & 3 \\
         {} & {} & {} & 3 & 117 & 507 & 275 & 86 & 15 \\
         {} & {} & {} & 4 & 14 & 115 & 194 & 467 & 210 \\
         {} & {} & {} & 5 & 9 & 19 & 23 & 195 & 754 \\
    \bottomrule
    \end{tabular}
    \vspace{2pt}
    \caption{Confusion matrix for error analysis.}
    \vspace{-7pt}
    \label{tab:cm}
\end{wraptable}

\subsection{Error Analysis}
%\tbc{dadi adds confusion matrix for filter aug's limitation}
In this section, we use the confusion matrix to analyze the limitations of augmented data. Specifically, we use 30,000 augmented data generated by Llama-3\textsubscript{8B} for the \textit{Arts} domain to fine-tune an SLM, then test it on 5,000 private training data and calculate the confusion matrix for 5 categories. 
The confusion matrix is shown in Table~\ref{tab:cm}. The number at index \textit{i,j} represents the count of samples where the true label is \textit{i} and the predicted value is \textit{j}.

From the table, we can observe that among the misclassified data, the model is most inclined to categorize neutral reviews whose label is 3, as 2 (2 indicates that the review is weakly negative). Subsequently, it tends to misclassify data points that belong to rating 2 as 1.
We go through the augmented data and find that the data labeled as 2 and 3 tends to contain both positive and negative opinions. 
Therefore, we suspect that SLMs fine-tuned solely on the augmented data are overly sensitive to the negative aspects of the reviews.

\subsection{Examples of Generated Data}
In this section, we conduct case studies to compare the generated data with original private data. We select two samples from the Health domain for synthetic data $D'$ with different privacy budgets, \textit{Gen KT} data $D^g$, FDKT's augmented data $D^a$ and private data $D$ to compare their data quality intuitively.

Table~\ref{tab:cases} lists a few representative cases for each data source.
For synthetic data, with a small $\epsilon = 1$, despite strict privacy protection, the generated reviews are contradictory and may not align with the corresponding ratings.
By increasing $\epsilon$, obvious improvements in the synthetic data quality can be observed.
In terms of Gen KT's data $D^g$, we can easily observe that the generated negative reviews frequently contain repetitive phrases like ``I’m extremely disappointed with my experience at this health
business.'' and ``I wouldn’t recommend this.''
Such repetitions imply that reviews augmented based only on the label information suffer from a lack of diversity and result in poor quality.
Instead, our FDKT's augmented data $D^a$ not only improves the quality of synthetic data $D'$ but also exhibits an increased data diversity due to the given in-context examples.
Still, if we compare $D^a$ with the client's private data $D$, it is evident that reviews in $D$ are longer and more descriptive than reviews in $D^a$.
Consequently, FDKT's augmented data quality is still inferior to that of the original private data.
This observation on cases intuitively explains why SLMs fine-tuned on $D^a$ underperform SLMs fine-tuned on $D$, even though $D^a$ has a much larger size.

\section{Ablation Studies}
\label{app:ablation}

{
\setlength{\tabcolsep}{6pt} % Default is 6pt
\begin{table*}[t]
\centering
%\small
%\fontsize{9pt}{9pt}\selectfont
%\setlength\extrarowheight{1pt}

  \begin{tabular}{l l c| cc}
    \toprule
    %{} & {} & {} & {} & {} & {} & {GARAGE}
    %   \\
       \multirow{2}{*}{\textbf{Method}} & 
       \multirow{2}{*}{\textbf{FT Data}} &
       \multirow{2}{*}{\textbf{$\epsilon$}} &
       \multicolumn{2}{c}{\textbf{Health}} \\
       {}  & {}  & {} & {Exact (\%)} & {Rough (\%)} \\
      \midrule

    %%%%% $D'$ + $D$
    %Syn FT & $D'$ & xx.xx & xx.xx & xx.xx & xx.xx & xx.xx & xx.xx\\
    
    %%% $\mathcal{F}(D')  + D$ 
    %Syn FT+$\mathcal{F}$ & $\mathcal{F}(D')$ & xx.xx & xx.xx & xx.xx & xx.xx & xx.xx & xx.xx\\

    Local FT & $D$ & - & 55.82 & 81.30\\
    
    FDKT & $D^a$ + $D$ & 1 & 57.30 & 81.90\\

    FDKT & $D^a$ + $D$ & 4 & 59.90 & 81.50\\
    
    FDKT & $D^a$ + $D$ & 8 & 57.20 & 81.80\\

    FDKT & $D^a$ + $D$ & 32& 58.30 & 81.10\\
    
    %FDKT & $D^a$ + $D$ & 64& xx.xx & xx.xx\\

    FDKT & $D^a$ + $D$ &256& 59.00 & 82.20\\

    \bottomrule
  \end{tabular}
\vspace{-0.05in}
\caption{\label{app: epsilon}
Evaluation of FDKT's performance with varied privacy budget $\epsilon$.
}

\vspace{-0.1in}
\end{table*}
}

{
\setlength{\tabcolsep}{6pt} % Default is 6pt
\begin{table*}[t]
\centering
%\small
%\fontsize{9pt}{9pt}\selectfont
%\setlength\extrarowheight{1pt}

  \begin{tabular}{l l c| cc}
    \toprule
    %{} & {} & {} & {} & {} & {} & {GARAGE}
    %   \\
       \multirow{2}{*}{\textbf{Method}} & 
       \multirow{2}{*}{\textbf{FT Data}} &
       \multirow{2}{*}{\textbf{Augmented Data \#}} &
       \multicolumn{2}{c}{\textbf{Shopping}} \\
       {}  & {}  & {} & {Exact (\%)} & {Rough (\%)} \\
      \midrule

    %%%%% $D'$ + $D$
    %Syn FT & $D'$ & xx.xx & xx.xx & xx.xx & xx.xx & xx.xx & xx.xx\\
    
    %%% $\mathcal{F}(D')  + D$ 
    %Syn FT+$\mathcal{F}$ & $\mathcal{F}(D')$ & xx.xx & xx.xx & xx.xx & xx.xx & xx.xx & xx.xx\\

    Local FT & $D$ & 0 & 50.08 & 70.30\\
    
    FDKT & $D^a$ + $D$ & 1,000  & 50.00 & 67.60\\

    FDKT & $D^a$ + $D$ & 5,000  & 58.00 & 76.20\\
    
    FDKT & $D^a$ + $D$ & 1,0000 & 59.50 & 79.60\\

    FDKT & $D^a$ + $D$ & 2,0000 & 58.20 & 80.40\\
    
    %FDKT & $D^a$ + $D$ & 64& xx.xx & xx.xx\\

    FDKT & $D^a$ + $D$ & 3,0000 & 56.13 & 78.43\\

    \bottomrule
  \end{tabular}
\vspace{-0.05in}
\caption{\label{app: aug_num}
Evaluation of FDKT's performance with a varied number of augmented data.
}

\vspace{-0.1in}
\end{table*}
}
%app: epsilon
\subsection{FDKT with Different Privacy Budgets}
In this section, we study privacy budgets' influence on FDKT's performance. 
Table~\ref{app: epsilon} lists FDKT's performance with $\epsilon$ = 1, 4, 8, 32, 256, where $\epsilon$ = 1 indicates the strictest privacy protection and $\epsilon$ = 256 fails to provide meaningful protection.
The results suggest that under small $\epsilon$, FDKT leads to similar evaluation performance.
For example, when $\epsilon$ = 4, FDKT's performance is even better than FDKT's results with $\epsilon$ = 8, 32.
Though strict privacy budgets compromise the synthetic data quality, our in-context augmented data can rectify the errors injected by DP.

%%%app: aug_num
\subsection{FDKT with Varied Numbers of Augmented Data}
Following the experimental settings mentioned in Section~\ref{sec:exp setup}, we control the number of augmented data to study FDKT's performance under varied $|D^a|$ on the Shopping domain.

Table~\ref{app: aug_num} illustrates that the performance of FDKT begins to improve and then diminishes as the volume of augmented data ( $|D^a|$ ) increases. This phenomenon suggests that simply increasing the quantity of augmented data is not the optimal strategy. Therefore, clients can choose a reasonable amount of augmented data to download from the server. For instance, 10,000 units of data may be sufficient for shopping data.

% In Table~\ref{app: aug_num}, the results further substantiate FDKT's effectiveness over \textit{Local FT}. 
% However, as the volume of augmented data $|D^a|$ increases, FDKT's performance fails to increase accordingly.
% Despite augmenting more data leads to high costs for clients, it does not necessarily improve local SLMs' performance.
% Therefore, our reported FDKT's performance is not optimized for the main experiments.
% If we consider economic factors, clients need to pay more with increased data augmentation.
% Consequently, it remains challenging to perform the multi-objective optimization by considering the privacy budget, economic factors and SLMs' utility.

\begin{table}[t]
    \centering
    \begin{tabular}{p{3cm} | p{10cm}}
    \toprule
    {Tasks} & {Prompts} \\
    \midrule
    Data Augmentation & \begin{minipage}[t]{10cm}
    \vspace{0pt} % Ensures top alignment
I will give you some customer feedback on \{'sub\_domain'\} related purchases. These reviews gradually shift from negative to positive from 1 star to 5 stars. 1 star represents the worst, 2 stars are better than 1 star, but still indicate a negative review. 3 stars represent a neutral review. 4 stars indicate a positive review, but less positive than 5 stars. 5 stars represent perfection. \\
    Please generate more similar samples for each rating star as shown in the following format, bearing in mind that the generated results should not copy or resemble the examples, and should be \{'sub\_domain'\}-related and align with the rating stars. The examples are delimited by '******': \\
    ******\\
    - <Rating>: \{'label\_1'\} stars. \\
    - <Review>: \{'input\_1'\}. \textbackslash n\textbackslash n \\
    ******\\
    - <Rating>: \{'label\_2'\} stars. \\
    - <Review>: \{'input\_2'\}. \textbackslash n\textbackslash n \\
    ...\\
    \end{minipage}
    \\
    \hline
    Data Filtering & \begin{minipage}[t]{10cm}
    \vspace{0pt} % Ensures top alignment
I will give you some customer review text samples with their rating stars, these samples are indexed starting from 0, the samples are delimited by '******':
\\
    ******\\
    0.\\
    - <Rating>: \{'label\_1'\} stars. \\
    - <Review>: \{'input\_1'\}. \textbackslash n\textbackslash n \\
    ******\\
    1.\\
    - <Rating>: \{'label\_2'\} stars. \\
    - <Review>: \{'input\_2'\}. \textbackslash n\textbackslash n \\
    ...\\
    These reviews gradually shift from negative to positive from 1 star to 5 stars. 1 star represents the worst, 2 stars are better than 1 star, but still indicate a negative review. 3 stars represent a neutral review. 4 stars indicate a positive review, but less positive than 5 stars. 5 stars represent perfection. Please filter out text that does not belong to customer reviews or does not meet the rating stars, and leave review texts that meet the labels. You should also filter out text that is too similar to other samples and keep the most representative ones. Your answer should begin with 'The eligible samples:\textbackslash n\textbackslash n' and the indexes of the texts you choose, use spaces to separate the indexes and do not provide duplicate indices or indices that exceed the maximum index of samples.
\\
\end{minipage}
    \\

\bottomrule
         
    \end{tabular}
    \caption{Prompt templates used for our experiments.
    }
    \vspace{-3mm}
    \label{tab:prompt}
\end{table}

\begin{table}[t]
    \centering
    \begin{tabular}{p{2.5cm} | p{10.5cm}}
    \toprule
    {Example Source} & {Content} \\
    \midrule
    Synthetic data $D'$ with $\epsilon = 1$  & \begin{minipage}[t]{10.5cm}
    %\vspace{0pt} % Ensures top alignment
    (1) If you are a doctor you can do this. It is very difficult, and if you can't you should not try this. When it starts to feel like a job, you can leave. (Rating: 1 star)\\
    \\%% line sep
    (2) The first day I got my appointment I felt so bad that my first visit was cancelled. The front desk staff is very friendly and helpful. My initial concerns were about the appointment being cancelled after the office visit.The doctor is very nice and attentive.I feel my first appointment was a waste of my time. (Rating: 1 star)\\
    \end{minipage}
    \\
    \hline
    Synthetic data $D'$ with $\epsilon = 32$ & \begin{minipage}[t]{10.5cm}
    %\vspace{0pt} % Ensures top alignment
    (1) Dr. David was wonderful with my children's needs. He took an interest in their mental health, and was a great help to me and my children.  I can't wait for our next visit! (Rating: 5 stars)\\
    \\%% line sep
    (2) The staff is so nice.  They are always friendly and attentive and the place is clean and safe.  I was worried about a little something but everything was good. (Rating: 3 stars)
\\
\end{minipage}
\\
    \hline
Gen KT data $D^g$ & \begin{minipage}[t]{10.5cm}
    %\vspace{0pt} % Ensures top alignment
    (1) I was extremely disappointed with my experience at this health business. The staff seemed overwhelmed and unfriendly, and the equipment was outdated and in disrepair. I left feeling frustrated and unsure if I would ever return. Overall, I would not recommend this business to anyone looking for a positive and effective health experience. (Rating: 1 star)\\
    \\%% line sep
    (2) I'm extremely disappointed with my experience at this health business. As someone who's trying to make positive lifestyle changes, I was expecting personalized guidance and support. Instead, I felt like I was just another number in a crowded room. The trainer was unprofessional and seemed put off when I asked questions. The facilities were also dirty and outdated. I wouldn't recommend this place to anyone. (Rating: 1 star)
\\
\end{minipage}
    \\
    \hline
Augmented data $D^a$ & \begin{minipage}[t]{10.5cm}
    %\vspace{0pt} % Ensures top alignment
    (1) Dr. David was wonderful with my children's needs. He took an interest in their mental health, and was a great help to me and my children.  I can't wait for our next visit! (Rating: 5 stars)\\
    \\%% line sep
    (2) I had a mixed experience at this dentist office. The hygienist was nice and did a good job cleaning my teeth, but the dentist was rushed and didn't take the time to explain what he was doing. The office was also quite noisy and chaotic. (Rating: 2 stars)
\end{minipage}
\\
    \\
    \hline
Private data $D$ & \begin{minipage}[t]{10.5cm}
    %\vspace{0pt} % Ensures top alignment
    (1) If you want to relax and enjoy some "me" time this is the place to go! If you are ok with being naked in front of other people then this is the place to go... there are multiple soaking tubs with different temps. The massage is great as well as the body scrub! I loved being able to relax in the sauna and other amenities that were included while I waited for my massage. (Rating: 4 stars)\\
    \\%% line sep
    (2) I love their food that you can get to go. The food is not properly labeled as far as how much it will cost. I guess it comes with the territory. it doesn't help that most of them do not speak any English so that is hard for somebody that doesn't speak complete Spanish. I would probably come here more if it seemed like they were customer-friendly. They are there to do a job and get it done. (Rating: 3 stars)
\end{minipage}
\\
\bottomrule
         
    \end{tabular}
    \caption{Case studies of synthetic data, augmented data and original private data on the Health domain.
    }
    \vspace{-3mm}
    \label{tab:cases}
\end{table}

%\clearpage
%\input{7-checklist}
\end{document}